\documentclass[onecolumn,11pt,oneside,draftclsnofoot]{IEEEtran}%

\usepackage{amsbsy}
\usepackage{floatflt} 

\usepackage{amsmath}
\usepackage{amssymb}
\usepackage{times}
\usepackage{graphicx}
\usepackage{xspace}
\usepackage{paralist} 
\usepackage{setspace} 
\usepackage{xypic}
\xyoption{curve}
\usepackage{latexsym}
\usepackage{theorem}
\usepackage{ifthen}
\usepackage{subfigure}

\usepackage{booktabs}
{\theoremheaderfont{\it} \theorembodyfont{\rmfamily}
\newtheorem{theorem}{Theorem}
\newtheorem{lemma}[theorem]{Lemma}
\newtheorem{definition}[theorem]{Definition}

\newtheorem{corollary}[theorem]{Corollary}

}

\def\ln{{\rm ln}}

\title{Sensor Networks with Random Links: Topology Design for Distributed Consensus}
\author{Soummya Kar and Jos\'e M.~F.~Moura$^{*}$
\thanks{The authors are with the Department of Electrical and Computer Engineering,
Carnegie Mellon University, Pittsburgh, PA, USA 15213 (e-mail:
soummyak@andrew.cmu.edu, moura@ece.cmu.edu, ph: (412)268-6341,
fax: (412)268-3890.)}
\thanks{Work supported by the DARPA DSO Advanced Computing and Mathematics Program
Integrated Sensing and Processing (ISP) Initiative under  ARO
grant \#~DAAD19-02-1-0180, by NSF under grants \#~ECS-0225449
and~\#~CNS-0428404, and by an IBM Faculty Award.}}

\begin{document}
\maketitle \thispagestyle{empty} \maketitle

\begin{abstract}
In a sensor network, in practice, the communication among sensors is subject to: \begin{inparaenum}[(1)] \item errors or failures at random times; \item  costs; and \item constraints since sensors and networks operate under scarce resources, such as  power, data rate, or communication. \end{inparaenum} The signal-to-noise ratio~(SNR) is usually a main factor in determining the probability of error (or of communication failure) in a link. These probabilities are then a proxy for  the SNR under which the links operate. The paper studies the problem of designing the topology, i.e.,   assigning the probabilities of reliable communication among sensors (or of link failures) to maximize the rate of convergence of average consensus, when the link communication costs are taken into account, and there is an overall communication budget constraint. To consider this problem,  we address a number of preliminary issues: \begin{inparaenum}[(1)] \item model the network as a random topology; \item  establish necessary and sufficient conditions  for mean
square sense (mss) and almost sure (a.s.) convergence of average consensus when network links fail; and, in particular, \item show
that a necessary and sufficient condition for both mss and a.s.~convergence is for the algebraic connectivity of the mean graph describing the network topology to be strictly positive. \end{inparaenum}   With these results, we formulate topology design, subject to random link failures and to a communication cost constraint, as a constrained convex optimization problem to which we apply semidefinite programming techniques. We show by an extensive numerical study that the optimal design improves significantly the convergence speed of the consensus algorithm and can achieve the asymptotic performance of a non-random network at a fraction of the communication cost.

%

\end{abstract}


\newpage

\section{Introduction}
\label{Introduction}
We consider the design of the optimal topology, i.e., the communication configuration of a sensor network that maximizes  the convergence rate of average consensus. Average consensus is a distributed algorithm that has been considered by Tsitsiklis in his PhD thesis, \cite{tsitsiklisphd84}, see also~\cite{tsitsiklisbertsekasathans86},  found application recently in several areas, and is the subject of active research, e.g,, \cite{jadbabailinmorse03,Olfati-2003,SensNets:Xiao05,blondelhendrickxolshevskytsitsiklis05}.

This topology design for sensor networks  has not received much attention in the literature. References~\cite{Olfati-ConsSmallWorld}
and~\cite{SensNets:AldosariAsilomar05} consider  restrict it to classes of random graphs,  in
particular, small-world topologies.
The more general question of designing the topology that maximizes the convergence rate, under a constraint on the
number of network links,  was considered in our previous work, \cite{Allerton06-K-M,Asilomar06-K-M,tsp06-K-A-M}, where we reduced to average consensus the problem of distributed inference in sensor networks; see also~\cite{Olfati-ramanujan}.

%
%
%
%
%
%
%
%

Realistic networks operate under stress: \begin{inparaenum}[(1)] \item noise and errors cause links to fail at random times; \item communication among sensors entails  a cost;  and \item  scarcity of resources constrain sensors and networks operation. \end{inparaenum} 
  We model such a non-deterministic network topology as a random
field. Specifically, we assume the following:
\begin{inparaenum}[1)] \item at each iteration
of the consensus algorithm, a network link is active with some
probability, referred to as link formation or utilization probability; \item
network links have different link formation probabilities;
 \item links fail or are alive independently of each other; and \item the link
formation probabilities remain constant across iterations.
\end{inparaenum}
Designing the network topology corresponds then to \begin{inparaenum}[(1)] \item fixing the probability, or fraction of time, each link is used, \item knowing that communication among  sensors may be cheap (e.g., sensors are geographically close), or expensive, and  \item recognizing that there is an overall budget constraint taxing the communication in the network. \end{inparaenum}

The paper extends our preliminary  convergence results, \cite{ICASSP07-K-M}, on networks
with random links. The recent paper~\cite{Jadbabai} adopts a similar model and  analyzes
convergence properties using ergodicity of stochastic matrices. 
Consensus with a randomized network also relates to gossip algorithms, \cite{Boyd-GossipInfTheory}, where only a single pair of randomly selected sensors is allowed to communicate at each iteration, and the communication exchanged by the nodes is averaged. In our randomized consensus, we use multiple  randomly selected links at each iteration and, in contradistinction with~\cite{Boyd-GossipInfTheory}, we design the optimal topology, i.e., the optimal weight (not simple average) and the optimal probabilities of edge utilization, recognizing that communication entails costs, and that there is a communication cost constraint.
 Other recent work on evolving topologies
 includes~\cite{SensNets:Olfati04} that considers continuous time
consensus in networks with switching topologies and communication
delays, and~\cite{Mesbahi} that studies distributed consensus when the
network is  a complete graph with identical link failure
probabilities on all links.



We outline the paper.
Section~\ref{ConsOverview,ProbForm} summarizes spectral
graph theory concepts like  the graph Laplacian~$L$ and the graph algebraic connectivity  $\lambda_2(L)$. 
  The Section formulates the problem of distributed average consensus with
random link failures. Sections~\ref{section:preliminaryresults} and
\ref{section:convergence} derive necessary and sufficient conditions  for convergence of
the mean state, mss convergence, and a.s.~convergence in terms of the average $\mbox{E}\left\{\lambda_2\left(L\right)\right\}$ and in terms of $\lambda_{2}\left(\overline{L}\right)$, where $\overline{L}=\mbox{E}\left(L\right)$.
Section~\ref{RateofConvergence} presents bounds on the mss
convergence rate. Section~\ref{ProbFormCommCost} addresses the topology design for random networks with communication cost constraints. We formulate a first version of the problem,
the randomized distributed consensus with a communication
cost constraint (RCCC), and then an alternate version, which we show is a convex constrained optimization problem,  to which we apply semidefinite programming~(SDP) techniques. Section~\ref{ARCCC_Perf_Num}  studies the performance of the topologies found by solving numerically the SDP optimization. We show that these designs can improve significantly the convergence rate, for example, by a factor of~$3$, when compared to geometric networks (networks where sensors communicate with every other sensor within a fixed radius) and that they can achieve practically the (asymptotic) performance of a nonrandom network at a fraction, e.g., 50~\%, of the communication cost per iteration. Section~\ref{Conclusions} concludes the
paper.

\section{Distributed Average Consensus}
\label{ConsOverview,ProbForm}
Subsection~\ref{subsection:nonrandomrandom}  presents two network models:
\begin{inparaenum}[Model 1)] \item \emph{Nonrandom} topology in
paragraph~\ref{subsubsection:nonrandom}; and \item \emph{Random}
topology in paragraph~\ref{subsubsection:random}.\end{inparaenum}
 \ Subsection~\ref{subsection:averageconsensus} considers
distributed average consensus with
\emph{nonrandom topologies} in
Paragraph~\ref{subsubsection:consensusnonrandom} and \emph{random}
topologies in Paragraph~\ref{subsubsection:averageconsensusrandom}.
We assume synchronous communication
throughout.

\subsection{Nonrandom and Random Topologies}
\label{subsection:nonrandomrandom} In a nonrandom topology,
the communication channels  stay
available whenever the sensors need to communicate. This model is
described in paragraph~\ref{subsubsection:nonrandom}, where we
recall  basic concepts from graph theory. In many sensor
network applications, it makes sense to consider that links among
sensors may fail or become alive at random times. This models,
for example, applications when the network uses an ARQ protocol and
no acknowledgement packet is received within the protocol time
window, in which case the transmitted packet is assumed to be
dropped or lost. This is also the case, when the transmission is
detected in error. The random topology introduced in
paragraph~\ref{subsubsection:random} models these networks.

\subsubsection{Nonrandom topology}
\label{subsubsection:nonrandom}
The nonrandom topology is defined by an undirected graph
$G=(V,\mathcal{E})$, where $V$ is the set of vertices that model
the sensors and $\mathcal{E}$ is the set of edges that model the
communication channels. We refer to~$G$ as the supergraph,
$\mathcal{E}$ as the \emph{superset} of edges, and edges in
$\mathcal{E}$ as \emph{realizable} edges or links. This
terminology becomes better motivated when we consider the random
topology in Subsection~\ref{subsubsection:random}. The
cardinalities of the sets $|V|=N$ and $|\mathcal{E}|=M$ give the
number of network sensors and the number of channels or links,
respectively. For the complete graph $G=(V,\mathcal{M})$, $\mathcal{M}$ is the set of all possible $N(N-1)/2$ edges. In practice, we are interested in sparse graphs,
i.e., $M\ll N(N-1)/2$. We label a node or vertex by an integer $n$,
where $n\in \{1,...,N\}$. Sensors~$n$ and~$l$ communicate if there
is an edge $(n,l)\in\mathcal{E}$. Since the graph is undirected,
if~$n$ communicates with~$l$, then~$l$ communicates with~$n$.  The
graph is called simple if it is devoid of loops (self-edges) and
multiple edges. It is connected if every vertex can be reached
from any other vertex, which in network terms may require a
routing protocol. The number $d_n$ of edges connected to
vertex~$n$ is called the degree of the vertex. A graph is regular
if every vertex has the same degree~$d$. Unless otherwise stated,
we consider only simple, connected graphs.

Associated with the graph~$G$ is its $N\times N$ adjacency
matrix~$\mathcal{A}$
\begin{equation}
\label{A_def} \mathcal{A}_{nl} = \left\{ \begin{array}{ll}
                    1 & \mbox{if $(n,l) \in \mathcal{E}$} \\
                    0 & \mbox{otherwise}
                   \end{array}
          \right.
\end{equation}
The neighborhood structure of the graph is defined by
\begin{equation}
\label{neighborhood} \forall 1\leq n\leq N:\:\:\Omega _{n} =
\left\{l\in V:~(n,l)\in \mathcal{E}\right\}
\end{equation}
The degree of node~$n$ is also the cardinality of its neighborhood
set
\begin{equation}
\label{deg_def} \forall 1\leq n\leq N:\:\:\mbox{d}_{n}=|\Omega
_{n}|
\end{equation}
Let $\mathcal{D} = \mbox{diag}(\mbox{d}_{1},...,\mbox{d}_{N})$ be
the degree matrix. The graph Laplacian matrix~$\mathcal{L}$ is
defined as
\begin{equation}
\label{L_def} \mathcal{L} = \mathcal{D}-\mathcal{A}
\end{equation}

The Laplacian~$\mathcal{L}$ is a symmetric positive-semidefinite
matrix; hence, all its eigenvalues are non-negative. We order the
Laplacian eigenvalues as
\begin{equation}
\label{lambdamathcalL}
0=\lambda_{1}(\mathcal{L})\leq\lambda_{2}(\mathcal{L})\leq
\cdots\leq\lambda_{N}(\mathcal{L})
\end{equation}
The multiplicity of the zero eigenvalue of the Laplacian is equal to
the number of connected components of the graph. Thus, for a
connected graph, $\lambda_{2}(\mathcal{L})>0$. In the literature,
$\lambda_{2}(\mathcal{L})$ is referred to as the algebraic
connectivity (or Fiedler value) of the network
(see~\cite{SensNets:Fiedler73}.) The normalized eigenvector
$\mathbf{u}_1(\mathcal{L})$ corresponding to the zero eigenvalue is
the normalized vector of ones
\begin{equation}
\label{eq:vector1}
\mathbf{u}_1\left(\mathcal{L}\right)=\frac{1}{\sqrt{N}}\mathbf{1}=\left[\frac{1}{\sqrt{N}}\cdots\frac{1}{\sqrt{N}}\right]^T
\end{equation}
For additional concepts from graph theory
see~~\cite{FanChung,Mohar,SensNets:Bollobas98}.

\subsubsection{Random Topology}
\label{subsubsection:random} We consider sensor networks where
failures may occur at random due to noise as when packets are
dropped. If a link fails at time~$i$, it can come back online at a
later time (a failed transmission may be succeeded by a successful
one.) We describe a graph model for this random topology. We start
with the model in paragraph~\ref{subsubsection:nonrandom} of a
simple, connected supergraph $G=\left(V,\mathcal{E}\right)$ with
$|V|=N$ and $|\mathcal{E}|=M$. The superset of edges $\mathcal{E}$
collects the realizable edges, i.e., the channels that are
established directly among sensors in the network when all
realizable links are online. These channels may fail at random
times, but if $(n,l)\notin\mathcal{E}$ then sensors~$n$ and~$l$ do
not communicate directly---of course, they still communicate by
rerouting their messages through one of the paths connecting them
in~$G$, since~$G$ is connected. We now construct the model for the
random topology problem, see also
\cite{ICASSP07-K-M,Jadbabai,Boyd-GossipInfTheory}.

To model this network with random link failures, we assume that the
 state, failed or online, of each link
$(n,l)\in\mathcal{E}$ over time $i=1,\cdots$ is a Bernoulli
process with probability of formation $P_{nl}$, i.e., the
probability of failure at time~$i$ is $1-P_{nl}$. We assume that
for any realizable edges $(n,l)\neq(m,k)$ the corresponding
Bernoulli processes are statistically independent. Under this
model, at each time~$i$, the the resulting topology is described
by a graph $G(i)=\left(V,E(i)\right)$. The edge set $E(i)$ and the
adjacency matrix~$A(i)$ are random, with $E(i)$ and $E(j)$, as
well as $A(i)$ and $A(j)$, statistically independent, identically
distributed (iid) for $i\neq j$.  Note that
$E(i)\subset\mathcal{E}$ and $\mathbf{0}\preceq
A(i)\preceq\mathcal{A}$, where $\mathbf{0}$ is the $N\times N$
zero matrix and $C\preceq D$ stands for $\forall 1\leq i,j\leq N:
C_{i,j}\leq D_{i,j}$. We can think of the set $E(i)$ as an
instantiation of a random binary valued $M$-tuple. The probability
of a particular instantiation $E(i)$ is
$\Pi_{(n,l)\in\mathcal{E}}P_{nl}$.
We collect the edge formation probabilities in the edge formation
probability matrix
\[
P=P^T=\left[P_{nl}\right], \:\:P_{n,n}=0
\]
 The diagonal elements are zero because the graph
is simple (no loops). The structure of~$P$ reflects the structure of
the adjacency matrix~$\mathcal{A}$ of the superset~$\mathcal{E}$,
i.e., $P_{nl}\neq0$ if and only if $\mathcal{A}_{nl}=1$. The
matrix~$P$ is not stochastic; its elements are $0\leq P_{nl}\leq 1$
but their row or column sums are not normalized to~$1$. Abusing
notation, we will refer to~$P$ as the probability distribution of
the~$E(i)$ and~$A(i)$.

We now consider the average consensus algorithm for both nonrandom
and random topologies.
\subsection{Average Consensus}
\label{subsection:averageconsensus} We overview average consensus,
see \cite{tsitsiklisphd84,tsitsiklisbertsekasathans86} and also for
recent work \cite{jadbabailinmorse03}. It computes by a distributed
algorithm the average of $x_n(0)$, $n=1,\cdots,N$ where $x_n(0)$ is
available at sensor~$n$. At time~$i$, each node exchanges its state
$x_n(i)$, $i=0,1,\cdots$ synchronously with its neighbors specified
by the graph edge neighborhood set, see eqn.~(\ref{neighborhood}).
In vector form, the $N$ states $x_n(i)$ are collected in the state
vector $\mathbf{x}(i)\in\mathbb{R}^{N\times 1}$. Define the average
$\overline{r}$ and the vector of averages
$\mathbf{x}_{\mbox{\scriptsize avg}}$
\begin{eqnarray}
\label{eqn:averager}
\overline{r}&=&\frac{1}{N}\mathbf{1}^{T}\mathbf{x}(0)\\
\label{def_xavg} \mathbf{x}_{\mbox{\scriptsize avg}} &=&\overline{r}\mathbf{1}\\
\label{def_xavg-b}
&=&\frac{1}{N}\mathbf{1}\mathbf{1}^{T}\mathbf{x}(0)\\
\label{def_xavg-c} &=&\frac{1}{N}J\mathbf{x}(0)
\end{eqnarray}
and where~$\mathbf{1}$ is the vector of ones,
see~(\ref{eq:vector1}), and $J=\mathbf{1}\mathbf{1}^{T}$. We next
consider the iterative average consensus  algorithm for both
nonrandom and random topologies.
\subsubsection{Average consensus: Nonrandom topology}
\label{subsubsection:consensusnonrandom} With the nonrandom
topology defined by the supergraph $G=\left(V,
\mathcal{E}\right)$, the state update by the average consensus
proceeds according to the iterative algorithm
\begin{eqnarray}
\label{lin_up_sens}
\forall i\geq0:\:\:\:x_{n}(i+1)&=&\mathcal{W}_{nn}x_{n}(i)+\sum_{l\in\Omega_{n}}\mathcal{W}_{nl}x_{l}(i)\\
\label{lin_up_mat}
\mathbf{x}(i+1)&=&\mathcal{W}\mathbf{x}(i)\end{eqnarray} where:
$\Omega_{n}$ is the neighborhood of sensor~$n$; $\mathbf{x}(i)$ is
the state vector collecting all states $x_n(i)$, $1\leq n\leq N$;
$\mathcal{W}_{nl}$ is the weight of edge $(n,l)$; and the matrix of
weights is $\mathcal{W}=\left[\mathcal{W}_{nl}\right]$. The sparsity
of~$W$ is determined by the underlying network connectivity, i.e.,
for $n\neq l$, the weight $\mathcal{W}_{nl}=0$ if $(n,l)\notin
{\mathcal{E}}$. Iterating~(\ref{lin_up_mat}),
\begin{eqnarray}
\label{prod_x-a}
\mathbf{x}(i) &=&\left(\prod_{j=0}^{i-1}\mathcal{W}\right)\mathbf{x}(0)\\
\label{prod_x-b} &=&\mathcal{W}^i \mathbf{x}(0)
\end{eqnarray}

A common choice for the weight matrix~$\mathcal{W}$ is the equal
weights matrix, \cite{SensNets:Xiao04},
\begin{equation}
\label{cons_W-a}
 \mathcal{W}=I-\alpha\mathcal{L}
\end{equation}
where $\mathcal{L}$ is the Laplacian associated with~$\mathcal{E}$,
and $\alpha\geq 0$ is a constant independent of time $i$. For the
equal weights matrix and a connected network, given the
ordering~(\ref{lambdamathcalL}) of the eigenvalues of~$\mathcal{L}$,
and that~$\alpha$ is nonnegative, the eigenvalues of~$\mathcal{W}$
can be reordered as
\begin{equation}\label{Eq:lambdaW}
    1=\lambda_1\left(\mathcal{W}\right)\geq\lambda_2\left(\mathcal{W}\right)\geq\cdots\geq\lambda_N\left(\mathcal{W}\right)
\end{equation}
The eigenvector corresponding to $\lambda_1\left(\mathcal{W}\right)$
is still the vector
$\mathbf{u}_1\left(\mathcal{W}\right)=\frac{1}{\sqrt{N}}\mathbf{1}$.



Reference~\cite{SensNets:Xiao04} studies the problem of optimizing
the nonzero weights~$\mathcal{W}_{nl}$ for maximizing convergence
rate when the adjacency matrix~$\mathcal{A}$ is known. In
particular, this reference shows that, for the equal weights case,
fastest convergence is obtained with
\begin{equation}\label{eq:equaloptimalweight}
    \alpha^{*}=\frac{2}{\lambda_2\left(\mathcal{L}\right)+\lambda_N\left(\mathcal{L}\right)}
\end{equation}
In~\cite{Allerton06-K-M,Asilomar06-K-M,tsp06-K-A-M}, we
consider this equal weight~$\mathcal{W}$ and  show that the class
of non-bipartite Ramanujan graphs provides the optimal (nonrandom)
topology under a constraint on the number of network links $M$,
see also~\cite{Olfati-ramanujan}. This optimality is in the
asymptotic limit of large~$N$, see the references for details.


\subsubsection{Average consensus: Random topology}
\label{subsubsection:averageconsensusrandom} At each time~$i$, the
graph $G(i)=\left(V,E(i)\right)$ is random. The distributed
average consensus algorithm still follows a vector iterative
equation like~(\ref{lin_up_mat}), except now the weight matrices
$W(i)$ are time dependent and random.
 We focus on the equal weights problem, 
\begin{equation}
\label{cons_W} W(i) = I-\alpha L(i)
\end{equation}
where $L(i)$ is the Laplacian of the random network at time~$i$.
The $L(i)$ are random iid matrices whose probability distribution
is determined by the edge formation probability matrix~$P$.
 Likewise, the weight matrices $W(i)$, $i=0,1,...$ are also iid random matrices.
We often drop the time
index~$i$ in the random matrices~$L(i)$ and~$W(i)$
or their statistics.
 Iterating~(\ref{lin_up_mat}) with this time dependent weight matrix leads to
\begin{equation}
\label{prod_x} \mathbf{x}(i) =
\left(\prod_{j=0}^{i-1}W(j)\right)\mathbf{x}(0)
\end{equation}

Since the weights $W_{nl}$ are random, the state $\mathbf{x}(i)$ is
also a random vector. Section~\ref{section:convergence} analyzes the
influence of the topology on the convergence properties as we
iterate~(\ref{prod_x}).


\section{Preliminary Results}
\label{section:preliminaryresults} Subsection~\ref{subsubsection:averageconsensusrandom} describes the random topology
model. The supergraph $G=\left(V,\mathcal{E}\right)$ is connected
and~$P$ is the matrix of edge formation probabilities. Since the $A(i)$, $L(i)$, and $W(i)$ are iid
\begin{eqnarray}
  \overline{A} &=& E\left[A(i)\right] \\
    \overline{L} &=& E\left[L(i)\right] \\
  \overline{W} &=& E\left[W(i)\right]\\
                &=&I-\alpha \overline{L}
\end{eqnarray}
i.e., their
means are time independent.
We establish properties of the Laplacian,
Subsection~\ref{subsection:laplacian}, and weight matrices,
Subsection~\ref{subsection:weight},  needed when studying the random
topology and random topology with communication cost constraint
problems in sections~\ref{section:convergence}
through~\ref{ProbFormCommCost}.

\subsection{Laplacian}
\label{subsection:laplacian} We list some properties of the mean
Laplacian and bound the expected value of the algebraic
connectivity of the random Laplacians by the algebraic
connectivity of the mean Laplacian.

\begin{lemma}
\label{lemma-.0} The mean adjacency matrix~$\overline{A}$ and mean
Laplacian are given by
\begin{eqnarray}\label{A_bar}
\overline{A}&=&P\\
\label{L_bar} \overline{L}_{nl}&=&\left\{\begin{array}{ll}
                    \sum_{m=1}^{N}P_{nm} & \mbox{if $n=l$} \\
                    -P_{nl} & \mbox{otherwise}
                    \end{array}
                    \right.
\end{eqnarray}
\end{lemma}
This Lemma is straightforward to prove. From the Lemma, it follows
that the mean adjacency matrix~$\overline{A}$ is not a $(0,1)$
matrix. Similarly. from the structure of the matrix
$\overline{L}$, see eqn.~(\ref{L_bar}), it follows that
$\overline{L}$ can be interpreted as the weighted Laplacian of a
graph $\overline{G}$ with non-negative link weights. In
particular, the weight of the link $(n,l)$ of $\overline{G}$ is
$P_{nl}$.
 The properties of the mean Laplacian  are similar to the properties of the Laplacian.
 We state them in the following two Lemmas.
\begin{lemma}
\label{lemma-.1} The mean Laplacian matrix
$\overline{L}=\mbox{E}\left[L(j)\right],~j=0,1,...$ is positive
semidefinite. Its eigenvalues can be arranged as
\begin{equation}
\label{eig_L_bar}
0=\lambda_{1}\left(\overline{L}\right)\leq\lambda_{2}\left(\overline{L}\right)\leq\cdots\leq\lambda_{N}\left(\overline{L}\right)
\end{equation}
where the normalized eigenvector associated with the zero
eigenvalue $\lambda_{1}\left(\overline{L}\right)$ is
\begin{equation}
\label{eq:u1Lbar}
\mathbf{u}_1\left(\overline{L}\right)=\frac{1}{\sqrt{N}}\mathbf{1}
\end{equation}
\end{lemma}
\begin{proof}
Let $\mathbf{z}\in\mathbb{R}^{N\times 1}$ be a non-zero vector.
Then, from eqn.~(\ref{L_bar}), we have
 \begin{equation}
\label{lemma-.1_1}
\mathbf{z}^{T}\overline{L}\mathbf{z}=\sum_{n,l}\overline{L}_{nl}z_{n}z_{l}=\frac{1}{2}\sum_{n\neq
l}P_{nl}(z_{n}-z_{l})^{2}
\end{equation}
Since the $P_{nl}$'s are non-negative,
$\overline{L}$ is positive semidefinite.
Eqn.(\ref{eq:u1Lbar}) follows readily from eqn.(\ref{lemma-.1_1}). 
\end{proof}

 Interpreting $\overline{L}$ as the
weighted Laplacian of the graph $\overline{G}$, we note that
$\lambda_{2}\left(\overline{L}\right)=0$ implies that $\overline{G}$
is not connected (see~\cite{Stefani, FanChung}.) In other words, if
$\lambda_{2}\left(\overline{L}\right)=0$, then $\overline{G}$ has at
least two disconnected components; hence, $\overline{L}$ takes the
form of a block diagonal matrix (after permuting the rows and
columns). Such matrices are called reducible matrices. Also, it
immediately follows (see~\cite{Stefani}) that, if $\overline{L}$ is
irreducible, then $\lambda_{2}\left(\overline{L}\right)\neq0$. Thus,
we get the following Lemma.

\begin{lemma}
\label{lemma:irreducible} Let the mean Laplacian be the weighted
Laplacian for a graph~$\overline{G}$.
\begin{equation}
\label{th3.2}
\lambda_{2}\left(\overline{L}\right)>0\Longleftrightarrow
\mbox{$\overline{L}$ is irreducible}\Longleftrightarrow
\overline{G}\:\mbox{is connected}
\end{equation}\end{lemma}

The convergence results in Section~\ref{Conv_mean} on the average
consensus involve the mean $\mbox{E}\left[\lambda_{2}(L)\right]$,
which is manifestly difficult to compute. A much easier quantity to
compute is $\lambda_{2}\left(\overline{L}\right)$. We relate here
the two. First, we show that $\lambda_{2}(L)$ is a concave function
of $L$.
\begin{lemma}
\label{lemma-6}$\lambda_{2}(L)$ is a concave function of $L$.
\end{lemma}
\begin{proof}
From the Courant-Fisher Theorem (see~\cite{FanChung, Mohar})
\begin{equation}
\label{l6.1}
\lambda_{2}(L)=\min_{\mathbf{z}\bot\mathbf{1}}\frac{\mathbf{z}^{T}L\mathbf{z}}{\mathbf{z}^{T}\mathbf{z}}
\end{equation}
Then for any two Laplacians $L_{1}$ and $L_{2}$ and $0\leq t\leq
1$ we have
\begin{eqnarray}
\label{l6.2} \lambda_{2}(tL_{1}+(1-t)L_{2}) & = &
\min_{\mathbf{z}\bot\mathbf{1}}\frac{\mathbf{z}^{T}(tL_{1}+(1-t)L_{2})\mathbf{z}}{\mathbf{z}^{T}\mathbf{z}}
\\ \nonumber & \geq &
t\min_{\mathbf{z}\bot\mathbf{1}}\frac{\mathbf{z}^{T}L_{1}\mathbf{z}}{\mathbf{z}^{T}\mathbf{z}}
+
(1-t)\min_{\mathbf{z}\bot\mathbf{1}}\frac{\mathbf{z}^{T}L_{2}\mathbf{z}}{\mathbf{z}^{T}\mathbf{z}}
\\ \nonumber & = & t\lambda_{2}(L_{1})+(1-t)\lambda_{2}(L_{2})
\end{eqnarray}
Thus $\lambda_{2}(L)$ is a concave function of $L$.
\end{proof}
\begin{lemma}
\label{lemma:lambda2concave}
\begin{equation}
\label{rel_mom}
\mbox{E}\left[\lambda_{2}(L)\right]\leq\lambda_{2}\left(\overline{L}\right)
\end{equation}
\end{lemma}
\begin{proof}
Follows from Lemma~\ref{lemma-6} and Jensen's inequality.
\end{proof}

\subsection{Weight matrices}
\label{subsection:weight} We consider properties of the (random and
mean) weight matrices. 
\begin{lemma}
\label{lemma:overlinewproperties} The eigenvalues of
$\overline{W}$ are
\begin{eqnarray}
\label{eig_W_bar}
1\leq j\leq N:&&\lambda_{j}\left(\overline{W}\right)=1-\alpha\lambda_{j}\left(\overline{L}\right)\\
\label{eig_W_bar_1} 1&=&\lambda_{1}\left(\overline{W}\right)\geq
\lambda_{2}\left(\overline{W}\right)\cdots\geq\lambda_{N}\left(\overline{W}\right)
\end{eqnarray}
The eigenvector corresponding to the eigenvalue
$\lambda_{1}\left(\overline{W}\right)$ is
\begin{equation}
\label{eq:u1overlineW}
\mathbf{u}_1\left(\overline{W}\right)=\frac{1}{\sqrt{N}}\mathbf{1}
\end{equation}
Similar results hold for~$W(i)$.
\end{lemma}
This Lemma follows immediately from the corresponding results on
the mean Laplacian and the~$L(i)$.

We now consider results on the spectral norm and its expected value
for the random matrices $W(i)$ and their mean~$\overline{W}$. These results are
used when studying convergence of the average consensus in
Section~\ref{section:convergence}.
\begin{lemma}
\label{lemma-.4} Let $\mathbf{z}\in\mathbb{R}^{N\times 1}$ and $\rho
(\cdot)$ be the spectral radius. Then
\begin{equation}
\label{lemma-.4_1} \forall \:W(j):\:\:
\left\|W(j)\mathbf{z}-\frac{1}{N}J\mathbf{z}\right\|\leq\rho\left
(W(j)-\frac{1}{N}J\right)\left\|\mathbf{z}-\frac{1}{N}J\mathbf{z}\right\|
\end{equation}
\end{lemma}
\begin{proof}
Decompose $W(j)$ through orthonormal eigenvectors as
$W(j)=U(j)\Lambda(j) U(j)^{T}$. From eqn.~(\ref{eig_W_bar_1}),
$\lambda_{1}(W(j))=1$ with normalized eigenvector
$\mathbf{u}_{1}(j)=\frac{1}{\sqrt{N}}\mathbf{1}$. Hence,
\begin{equation}
\label{l2.2}
\mathbf{z}=\frac{1}{N}J\mathbf{z}+\sum_{k=2}^{N}c_{k}(j)\mathbf{u}_{k}(j)
\end{equation}
where $c_{k}(j)=\mathbf{u}_{k}(j)^{T}\mathbf{z},~k=2,...,N$. Then
\begin{equation}
\label{l2.3} W(j)\mathbf{z}
=\frac{1}{N}J\mathbf{z}+\sum_{k=2}^{N}c_{k}(j)\lambda_{k}(W(j))\mathbf{u}_{k}(j)
\end{equation}
It follows that
\begin{eqnarray}
\label{l2.4} \left\| W(j)\mathbf{z}-\frac{1}{N}J\mathbf{z}\right\|
& = & \left\|
\sum_{k=2}^{N}c_{k}(j)\lambda_{k}(W(j))\mathbf{u}_{k}(j)\right\|\\
\nonumber & \leq & \rho \left(W(j)-\frac{1}{N}J\right)
\left\| \sum_{k=2}^{N}c_{k}(j)\mathbf{u}_{k}(j)\right\|\\
\nonumber & = & \rho \left(W(j)-\frac{1}{N}J\right)\left\|
\mathbf{z}-\frac{1}{N}J\mathbf{z}\right\|
\end{eqnarray}
This proves the Lemma.
\end{proof}

\begin{lemma}
\label{lemma-.2}  We have
\begin{eqnarray}
\label{def_rho} \rho\left(\overline{W}-\frac{1}{N}J\right)&=&\max
\left(|\lambda_{2}\left(\overline{W}\right)|,|\lambda_{N}\left(\overline{W}\right)|\right) = \max\left(\lambda_{2}\left(\overline{W}\right),-\lambda_{N}\left(\overline{W}\right)\right)\\
\label{def_rho-wi} \rho\left(W(i)-\frac{1}{N}J\right)&=&\max
\left(|\lambda_{2}\left(W(i)\right)|,|\lambda_{N}\left(W(i)\right)|\right)
=
\max\left(\lambda_{2}\left(W(i)\right),-\lambda_{N}\left(W(i)\right)\right)
\end{eqnarray}
\end{lemma}
\begin{proof}
We prove only the Lemma for~$\overline{W}$. Matrix $\frac{1}{N}J$ is
rank one, and the its non-zero eigenvalue is~$1$ with normalized
eigenvector $\frac{1}{\sqrt{N}}\mathbf{1}$. Hence, from
eqn.~(\ref{eig_W_bar_1}), the eigenvalues of
$\left(\overline{W}-\frac{1}{N}J\right)$ are~$0$ and
$\lambda_{2}\left(\overline{W}\right),...,\lambda_{N}\left(\overline{W}\right)$.
By the definition of spectral radius and eqn.~(\ref{eig_W_bar_1}),
\begin{equation}
\label{lemma-.2_1} \rho
\left(\overline{W}-\frac{1}{N}J\right)=\max
\left(0,|\lambda_{2}\left(\overline{W}\right)|,...,|\lambda_{N}\left(\overline{W}\right)|\right)=\max
\left(|\lambda_{2}\left(\overline{W}\right)|,|\lambda_{N}\left(\overline{W}\right)|\right)
\end{equation}
Also, noting that
$\lambda_{2}\left(\overline{W}\right)\geq\lambda_{N}\left(\overline{W}\right)$,
it follows from eqn.~(\ref{lemma-.2_1}) that
\begin{equation}
\label{def_rho-b} \rho
\left(\overline{W}-\frac{1}{N}J\right)=\max\left(\lambda_{2}\left(\overline{W}\right),-\lambda_{N}\left(\overline{W}\right)\right)
\end{equation}
\end{proof}

We now consider the convexity of the spectral norm as a function
of~$\alpha$ and~$L$. 
\begin{lemma}
\label{lemma:convexrho-W-JNalpha} For a given $L$, $\rho
\left(W-\frac{1}{N}J\right)$ is a convex function of~$\alpha$.
For a given $\alpha$, $\rho \left(W-\frac{1}{N}J\right)$ is a
convex function of~$L$.
\end{lemma}
\begin{proof}
We prove the convexity with respect to~$\alpha$ only. Let
$\alpha_{1},\alpha_{2}\in\mathbb{R}$ and $0\leq t\leq 1$. For
symmetric matrices the spectral radius is equal to the matrix
2-norm. We get
\begin{eqnarray}
\nonumber \rho
\left(I-\left(t\alpha_{1}+(1-t)\alpha_{2}\right)L-\frac{1}{N}J\right)
& = & \left\| I - t\alpha_{1}L-(1-t)\alpha_{2}L
-\frac{1}{N}J\right\|_{2} \\ \nonumber & = &
\left\|t\left(I-\alpha_{1}L-\frac{1}{N}J\right)+(1-t)\left(I-\alpha_{2}L-\frac{1}{N}J\right)\right\|_{2}\\
\nonumber & \leq &
\left\|t\left(I-\alpha_{1}L-\frac{1}{N}J\right)\right\|_{2}+
\left\|(1-t)\left(I-\alpha_{2}L-\frac{1}{N}J\right)\right\|_{2}\\
\label{l4.1} & = & t\rho \left(I-\alpha_{1}L-\frac{1}{N}J\right) +
(1-t)\rho \left(I-\alpha_{2}L-\frac{1}{N}J\right)
\end{eqnarray}
that proves the Lemma.
\end{proof}
The next Lemma considers the convexity of the expected value of
the spectral norm, taken over the probability distribution of the
Laplacian. The following Lemma bounds $\mbox{E}\left[\rho
\left(W-\frac{1}{N}J\right)\right]$.
\begin{lemma}
\label{lemma-4} For a given probability distribution (and hence
$P$) of $L$,
$\mbox{E}\left[\rho\left(W-\frac{1}{N}J\right)\right]$ is convex
on~$\alpha$.
\end{lemma}
\begin{proof}
The convexity of
$\mbox{E}\left[\rho\left(W-\frac{1}{N}J\right)\right]$  follows from
Lemma~\ref{lemma:convexrho-W-JNalpha}, eqn.~(\ref{l4.1}), and the
properties of Lebesgue integration.
\end{proof}


\begin{lemma}
\label{lemma-3} For a given choice of $\alpha$,
\begin{equation}
\label{l3.1} \mbox{E}\left[\rho
\left(W-\frac{1}{N}J\right)\right]\geq \rho
\left(\overline{W}-\frac{1}{N}J\right)
\end{equation}
\end{lemma}
\begin{proof}
 The Lemma follows from Lemma~\ref{lemma:convexrho-W-JNalpha} and Jensen's inequality.
\end{proof}

\section{Convergence of Average Consensus: Random Topology}
\label{section:convergence} For average consensus in random
topologies, we start by considering the convergence of the state
\begin{equation}
\label{conv_x} \forall\mathbf{x}(0)\in\mathbb{R}^{N\times
1}:\:\:\lim_{i\rightarrow\infty}\mathbf{x}(i)=\mathbf{x}_{\mbox{\scriptsize{avg}}}
\end{equation}
in some appropriate probabilistic sense.
Subsection~\ref{Conv_mean} studies convergence of the mean vector,
$\mbox{E}\left[\mathbf{x}(i)\right]$, Subsection~\ref{MeanSqConv}
considers convergence in the mean-square-sense~(mss), and almost
sure convergence (convergence with probability 1) is treated in
Subsection~\ref{AlmostSureConv}.

\subsection{Mean state convergence}
\label{Conv_mean} The sequence of expected state vectors converges
if
\begin{equation}
\label{convcond_Ex} \lim_{i\rightarrow\infty}\left\|
\mbox{E}\mathbf{x}(i)-\mathbf{x}_{\mbox{\scriptsize{avg}}}\right\|
= 0
\end{equation}
For simplicity, we assume $\|\cdot\|$ to be the
$\mathcal{L}_{2}$-norm. We analyze the convergence of the mean state
vector in ~\ref{subsubsection:meanstateconvergence} and then study
the topology that optimizes its convergence rate
in~\ref{subsubsection:meantopology}.
\subsubsection{Mean state convergence}
\label{subsubsection:meanstateconvergence} The mean state
evolution is given in the following Lemma.
\begin{lemma}
\label{lemma:Ex_avg}
Recall $\mathbf{x}_{\mbox{\scriptsize{avg}}}$  given
in~(\ref{def_xavg}). Then
\begin{equation}
\label{Ex_avg}
\mbox{E}\mathbf{x}(i)-\mathbf{x}_{\mbox{\scriptsize{avg}}}=\left(\overline{W}-\frac{1}{N}J\right)^{i}(\mathbf{x}(0)-\mathbf{x}_{\mbox{\scriptsize{avg}}})
\end{equation}
\end{lemma}
\begin{proof}

Using eqn.~(\ref{prod_x}) and the fact that the matrices~$W(i)$
are iid
\begin{equation}
\label{Ex}
\mbox{E}\left[\mathbf{x}(i)\right]=\overline{W}^{i}\mathbf{x}(0)
\end{equation}
The Lemma follows by recalling that $\mathbf{1}$ is an eigenvector
of $\overline{W}$.

\end{proof}
Convergence of the mean is now straightforward.

\begin{theorem}
\label{lemma-.3} A necessary and sufficient condition for the mean
to converge is 
\begin{equation}
\label{Ex_conv_cond} \rho \left(\overline{W}-\frac{1}{N}J\right)<1
\end{equation}
\end{theorem}
\begin{proof}
Lemma~\ref{lemma:Ex_avg} shows that the convergence of the mean is
equivalent to deterministic distributed average consensus. The
necessary and sufficient condition for convergence then follows
from references~\cite{tsp06-K-A-M,Boyd}.
\end{proof}
\subsubsection{Fastest mean convergence topology}
\label{subsubsection:meantopology} We introduce the definition of
convergence factor.
\begin{definition}[Mean convergence factor]
\label{def:convergencefactormean} If
$\rho\left(\overline{W}-\frac{1}{N}J\right)<1$, we call $\rho
\left(\overline{W}-\frac{1}{N}J\right)$ the mean convergence
factor of the consensus algorithm.
\end{definition}
For fastest mean convergence, $\rho
\left(\overline{W}-\frac{1}{N}J\right)$ should be as small as
possible. Hence, the optimal topology with respect to convergence of
the mean state vector is the topology that minimizes this
convergence factor. We address this problem in the following two
Theorems.

We note that $\rho (\overline{W}-\frac{1}{J}N)$ is a function of
both $\alpha$ and $\overline{L}$. In the following Theorem, we
state conditions on $\overline{L}$ that guarantee that we can
choose an $\alpha$ for which there is convergence of the mean.
\begin{theorem}
\label{theorem-1} A necessary condition for the mean to converge
is
\begin{equation}
\label{eq:necessarymean} \lambda_{2}\left(\overline{L}\right)>0
\end{equation}
A sufficient condition is~(\ref{eq:necessarymean}) and
\begin{equation}
\label{Ex_alpha_bound}0<\alpha<2/\lambda_{N}\left(\overline{L}\right)
\end{equation}

\end{theorem}
\begin{proof}
We first prove the necessary condition by contradiction. Let
$\lambda_{2}\left(\overline{L}\right)=0$. From
eqn.~(\ref{eig_W_bar}), it follows that
$\lambda_{2}\left(\overline{W}\right)=1$. Then, from
eqn.~(\ref{def_rho}), we have $\rho
\left(\overline{W}-\frac{1}{N}J\right)\geq 1$, for every choice of
$\alpha$. Hence, from Lemma~\ref{lemma-.3}, it follows that, if
$\lambda_{2}\left(\overline{L}\right)=0$, the mean vector does not
converge for any choice of $\alpha$. This proves the necessary
condition.

 For sufficiency, we assume that
$\lambda_{2}\left(\overline{L}\right)>0$. Then, generalizing the
results in~\cite{Boyd} to non-binary $(0-1)$ matrices, it can be
shown that
\[
\rho \left(\overline{W}-\frac{1}{N}J\right)<1
\:\:\mbox{\emph{iff}}\:\:
0<\alpha<2/\lambda_{N}\left(\overline{L}\right)
\]
 which then guarantees convergence of the mean state vector.
 \end{proof}

If $\lambda_{2}\left(\overline{L}\right)>0$,
Theorem~\ref{theorem-1} and eqn.~(\ref{Ex_alpha_bound}) give the
values of $\alpha$ that lead to the convergence of the mean vector
in terms of $\lambda_{N}\left(\overline{L}\right)$, a quantity
easily evaluated since $\overline{L}$ is given by
eqn.~(\ref{L_bar}).

The following Theorem gives the choice of $\alpha$ leading to the
fastest convergence of the mean.
\begin{theorem}
\label{lemma-1} Let $\lambda_{2}\left(\overline{L}\right)>0$. Then
the choice of $\alpha$ that minimizes $\rho
\left(\overline{W}-\frac{1}{N}J\right)$ and hence maximizes the
convergence rate of the mean state vector is
\begin{equation}
\label{eq:alphamax} \alpha^{\star}
=\frac{2}{\lambda_{2}\left(\overline{L}\right)+\lambda_{N}\left(\overline{L}\right)}
\end{equation}
The corresponding minimum $\rho(\cdot)$ is
\begin{equation}
\label{eq:rhomin} \rho_{\mbox{\scriptsize
min}}\left(\overline{W}-\frac{1}{N}J\right) =
\frac{1-\lambda_{2}\left(\overline{L}\right)/\lambda_{N}\left(\overline{L}\right)}{1+
\lambda_{2}\left(\overline{L}\right)/\lambda_{N}\left(\overline{L}\right)}
\end{equation}
\end{theorem}
\begin{proof}
It follows by generalizing the result in~\cite{Boyd} to
 non-binary matrices.
\end{proof}

This section derived necessary and sufficient conditions for the
convergence of the mean in terms of
$\lambda_{2}\left(\overline{L}\right)$. Also, it provided the
values of $\alpha$ that guarantee convergence when
$\lambda_{2}\left(\overline{L}\right)>0$. The next Subsection
considers mss convergence of average consensus.

\subsection{Mean Square Convergence}
\label{MeanSqConv} This Section studies mean-square convergence,
which implies convergence of the mean, but not the reverse. We say
that the algorithm converges in the mean-square sense~(mss) iff
\begin{equation}
\label{def_mss_conv} \forall\mathbf{x}(0)\in\mathbb{R}^{N\times
1}:\:\:\lim_{i\rightarrow\infty}
\mbox{E}\left\|\mathbf{x}(i)-\mathbf{x}_{\mbox{\scriptsize{avg}}}\right\|
= 0
\end{equation}
We need the following lemma first.
\begin{lemma}
\label{lemma-2} For any $\mathbf{x}(0)\in\mathbb{R}^{N\times 1}$
\begin{equation}
\label{l2.1}
\left\|\mathbf{x}(i+1)-\mathbf{x}_{\mbox{\scriptsize{avg}}}\right\|~\leq
\left(\prod_{j=0}^{i}\rho\left(W(j)-\frac{1}{N}J\right)\right)\left\|\mathbf{x}(0)-\mathbf{x}_{\mbox{\scriptsize{avg}}}\right\|
\end{equation}
\end{lemma}
\begin{proof}

We have
\begin{eqnarray}
\label{l2.5}
\left\|\mathbf{x}(i+1)-\mathbf{x}_{\mbox{\scriptsize{avg}}}\right\|
& = &
\left\|\left(\prod_{j=0}^{i}W(j)\right)\mathbf{x}(0)-\frac{1}{N}J\mathbf{x}(0)\right\|
\\ \nonumber
& = & \left\|
W(i)\left(\prod_{j=0}^{i-1}W(j)\mathbf{x}(0)\right)-\frac{1}{N}J\left(\prod_{j=0}^{i-1}W(j)\mathbf{x}(0)\right)\right\|
\end{eqnarray}
where we have used the fact that
\[
\frac{1}{N}J\left(\prod_{j=0}^{i-1}W(j)\mathbf{x}(0)\right)=\frac{1}{N}J\mathbf{x}(0)
\]
 From 
Lemma~\ref{lemma-.4}, it then follows
\begin{eqnarray}
\label{l2.10}
\left\|\mathbf{x}(i+1)-\mathbf{x}_{\mbox{\scriptsize{avg}}}\right\|
 & \leq & \rho\left(W(i)-\frac{1}{N}J\right)\left\|\left(\prod_{j=0}^{i-1}W(j)\mathbf{x}(0)\right)-
                    \frac{1}{N}J\left(\prod_{j=0}^{i-1}W(j)\mathbf{x}(0)\right)\right\|\\
\nonumber & = &
\rho\left(W(i)-\frac{1}{N}J\right)\left\|\mathbf{x}(i)-\mathbf{x}_{\mbox{\scriptsize{avg}}}\right\|
\end{eqnarray}
Repeating the same argument for $j=0 \:\mbox{to}\: i$ we finally
get
\begin{equation}
\label{l2.6}
\left\|\mathbf{x}(i+1)-\mathbf{x}_{\mbox{\scriptsize{avg}}}\right\|~\leq
\left(\prod_{j=0}^{i}\rho\left(W(j)-\frac{1}{N}J\right)\right)\left\|\mathbf{x}(0)-\mathbf{x}_{\mbox{\scriptsize{avg}}}\right\|
\end{equation}
This proves the Lemma.
\end{proof}

 The following
Theorem gives a sufficient condition for mss convergence.
\begin{theorem}
\label{theorem-2} If
$\mbox{E}\left[\rho\left(W-\frac{1}{N}J\right)\right]<1$, the
state vector sequence $\{\mathbf{x}(i)\}_{i=0}^{\infty}$ converges
in the mss 
\begin{equation}
\label{th2.1}
\lim_{i\rightarrow\infty}\mbox{E}\left\|\mathbf{x}(i)-\mathbf{x}_{\mbox{\scriptsize{avg}}}\right\|
= 0,~\forall\mathbf{x}(0)\in\mathbb{R}^{N\times 1}
\end{equation}
\end{theorem}
\begin{proof}
Taking expectation on both sides of eqn.~(\ref{l2.1}) in
Lemma~\ref{lemma-2} and using the iid of the $W(j)$'s
\begin{equation}
\label{th2.2} \mbox{E}\left\|
\mathbf{x}_{i}-\mathbf{x}_{\mbox{\scriptsize{avg}}}\right\|~\leq
\left(\mbox{E}\left[\rho
\left(W-\frac{1}{N}J\right)\right]\right)^{i-1}\left\|\mathbf{x}_{0}-\mathbf{x}_{\mbox{\scriptsize{avg}}}\right\|
\end{equation}
where we dropped the index~$i$ in~$W(i)$. The Theorem then
follows.
\end{proof}
Like the Definition~\ref{def:convergencefactormean} for mean
convergence factor, we introduce the mss convergence factor. First,  note that $\mbox{E}\left[\rho
\left(W-\frac{1}{N}J\right)\right]$ is a function of the weight
$\alpha$ and the probability of edge formation matrix $P$ (or
$\overline{L}$ from~(\ref{L_bar}).) 
\begin{definition}[mss convergence factor, mss convergence rate]
\label{def:mssconvergencefactor} If $\mbox{E}\left[\rho
\left(W-\frac{1}{N}J\right)\right]<1$, call
$C\left(\alpha,\overline{L}\right)$ and
$S_{\mbox{\scriptsize{g}}}(\alpha,\overline{L})$ the mss
convergence factor and the mss convergence gain per iteration (or
the mss convergence rate), respectively, where
\begin{eqnarray}
\label{eq:mssconvergencefactor}
C\left(\alpha,\overline{L}\right)&=&\mbox{E}\left[\rho
\left(W-\frac{1}{N}J\right)\right]\\
\label{def_Sg} S_{\mbox{\scriptsize{g}}}(\alpha,\overline{L})
&=&-\ln\,\,
C\left(\alpha,\overline{L}\right)\\
\label{def_Sg-b} &=& \ln\left(\frac{1}{\mbox{E}\left[\rho
\left(W-\frac{1}{N}J\right)\right]}\right)
\end{eqnarray}
\end{definition}

\begin{corollary}
\label{corollary:meanandmssconvergence}
 mss convergence cannot be faster than convergence of the mean vector.

\end{corollary}
The Corollary follows from the Theorem and Lemma~\ref{lemma-3}.

Theorem~\ref{theorem-2} shows that the smaller the mss~convergence
factor $C\left(\alpha,\overline{L}\right)=\mbox{E}\left[\rho
\left(W-\frac{1}{N}J\right)\right]$ is, the faster the
mss~convergence.  The actual value of
$C\left(\alpha,\overline{L}\right)$ depends both on the
probability distribution of the Laplacian~$L$ and the constant
weight~$\alpha$.
 However, the probability
distribution of~$L$ must satisfy certain conditions to guarantee
that there are values of~$\alpha$ that lead to mss convergence. Otherwise, no choice of~$\alpha$ will result in
mss~convergence. The next Theorem considers this issue. Before
stating the Theorem, let $d_{\mbox{\scriptsize{max}}}$ be the
maximum degree of the graph with edge set $E=\mathcal{E}$ and
define
\begin{equation}
\label{th3.6} \alpha_{\mbox{\scriptsize{mss}}} =
\frac{1}{2d_{\mbox{\scriptsize{max}}}}
\end{equation}
\begin{theorem}
\label{theorem-3} There is an $\alpha$ such that the consensus
algorithm converges in mss iff
$\lambda_{2}\left(\overline{L}\right)>0$. In other words, if
$\lambda_{2}\left(\overline{L}\right)>0$, we can find an $\alpha$,
in particular, $\alpha=\alpha_{\mbox{\scriptsize mss}}$ defined
in~(\ref{th3.6}), that leads to mss~convergence. If
$\lambda_{2}\left(\overline{L}\right)=0$, no choice of $\alpha$
will result in mss~convergence.
\end{theorem}
\begin{proof}
We first prove the sufficiency part. The proof is constructive, and
we show that, if $\lambda_{2}\left(\overline{L}\right)>0$, we can
find an $\alpha$ for which
\[
C\left(\alpha,\overline{L}\right)=\mbox{E}\left[\rho\left(W-\frac{J}{N}\right)\right]<1
\]
Convergence then follows from Theorem~\ref{theorem-2}.

 Let
$\lambda_{2}\left(\overline{L}\right)>0$. By
Lemma~\ref{lemma:irreducible}, $\overline{L}$ is irreducible. From
irreducibility of $\overline{L}$, with non-zero probability, we
have graph realizations for which~$L$ is irreducible and so
$\lambda_{2}(L)>0$. In particular, with non-zero probability, we
can have a realization for which the edge set $E=\mathcal{E}$; by
assumption, this network is irreducible and hence connected
(because the corresponding Laplacian matrix has the same sparsity
pattern of $\overline{L}$ with non-zero entries of $\overline{L}$
replaced by ones.) Hence, with non-zero probability,
$\lambda_{2}(L)>0$, which makes
$\mbox{E}\left[\lambda_{2}(L)\right]>0$. Thus we have
\begin{equation}
\label{th3.3}
\lambda_{2}\left(\overline{L}\right)>0\Longrightarrow
\mbox{E}\left[\lambda_{2}(L)\right]>0
\end{equation}

%
Let $d_{\mbox{\scriptsize{max}}}(G)$ be the maximum vertex degree of
graph $G$. Then, from spectral graph theory, see~\cite{Stefani},
\begin{equation}
\label{th3.4} \lambda_{N}(L(G))\leq
2d_{\mbox{\scriptsize{max}}}(G)
\end{equation}
We now claim mss convergence for
$\alpha=\alpha_{\mbox{\scriptsize mss}}$. 
  From Lemma~\ref{lemma-.2} and~(\ref{eig_W_bar}),
\begin{eqnarray}
\label{th3.7} \rho\left(W-\frac{1}{N}J\right) & = & \max
\left(\lambda_{2}(W),-\lambda_{N}(W)\right) \\ \nonumber & = &
\max \left(1-\alpha_{\mbox{\scriptsize mss}}\lambda_{2}(L),
\alpha_{\mbox{\scriptsize mss}}\lambda_{N}(L)-1\right) \\
\nonumber & = & 1-\alpha_{\mbox{\scriptsize mss}}\lambda_{2}(L)
\end{eqnarray}
where the last step follows from the fact that from eqn.~(\ref{th3.4}) and~(\ref{th3.6})
\begin{equation}
\label{th3.100} 1-\alpha_{\mbox{\scriptsize
mss}}\lambda_{2}(L)\geq 0\geq \alpha_{\mbox{\scriptsize
mss}}\lambda_{N}(L)-1
\end{equation}
Taking expectation on both sides of eqn.~(\ref{th3.7}), and since
$0<\mbox{E}\left[\lambda_{2}(L)\right]\leq
2d_{\mbox{\scriptsize{max}}}$, we get
\begin{eqnarray}
\label{th3.8}
C\left(\alpha,\overline{L}\right)&=&\mbox{E}\left[\rho\left(W-\frac{1}{N}J\right)\right]\\
\nonumber &=&1-\alpha_{\mbox{\scriptsize
mss}}\mbox{E}\left[\lambda_{2}(L)\right]\\
\nonumber &<& 1
\end{eqnarray}
mss convergence then follows from Theorem~\ref{theorem-1}.
This proves the sufficiency part.

The necessary condition follows from the fact that, if
$\lambda_{2}\left(\overline{L}\right)=0$, Theorem~\ref{theorem-1}
precludes convergence of the mean vector. Since, by
Corollary~\ref{corollary:meanandmssconvergence}, convergence of
the mean is necessary for mss convergence, we conclude
that, if $\lambda_{2}\left(\overline{L}\right)=0$, no choice of
$\alpha$ will result in mss convergence.
\end{proof}

Theorem~\ref{theorem-3} gives necessary and sufficient conditions
on the probability distribution of the Laplacian $L$ for mean
square convergence. This is significant as it relates mss
convergence to the network topology. Because this condition is in
terms of the algebraic connectivity of the mean Laplacian
associated with the probability distribution of edge
formation~$P$, it is straightforward to check.

\subsection{Almost Sure Convergence}
\label{AlmostSureConv}

We extend the results of the earlier sections and show that
$\lambda_{2}\left(\overline{L}\right)>0$ is also a necessary and
sufficient condition for a.s.~convergence of the
sequence $\left\{\mathbf{x}(i)\right\}_{i=0}^{\infty}$. Before
proceeding to a formal statement and proof of this, we recall some
basic facts about the convergence of (scalar) random variables.

\begin{definition}[A.S.~Convergence of random variables]
\label{def:asconv} Let
$\left\{\mathbf{\xi}_{i}\right\}_{i=0}^{\infty}$ be a sequence of
random variables defined on some common probability space
$(\Omega,\mathcal{F},\mathbb{P})$. Then
$\left\{\mathbf{\xi}_{i}\right\}_{i=0}^{\infty}$ converges a.s.~to another random variable $\mathbf{\xi}$ defined on
$(\Omega,\mathcal{F},\mathbb{P})$
$(\mathbf{\xi}_{i}\rightarrow\mathbf{\xi} \mbox{a.s.})$ if
\begin{equation}
\label{def:asconv1} \mathbb{P}\left(\omega\in\Omega ~:~
\mathbf{\xi}_{i}(\omega)\xrightarrow[i\rightarrow\infty]{}\mathbf{\xi}(\omega)\right)
= 1
\end{equation}
\end{definition}
This definition readily extends to random vectors, where a.s.~convergence means a.s.
~convergence of each component
(see~\cite{Kallenberg,Roussas}.)\\
We also recall that mss convergence of a sequence of random
variables $\left\{\mathbf{x}(i)\right\}_{i=0}^{\infty}$ implies
convergence in probability through Chebyshev's inequality. Also, we
note that convergence in probability implies a.s.~convergence of a
subsequence (see~\cite{Galambos,Roussas}.)\\
We now formalize the theorem for almost sure convergence of the
state vector sequence
$\left\{\mathbf{x}(i)\right\}_{i=0}^{\infty}$.
\begin{theorem}
\label{th:as} A necessary and sufficient condition for a.s.~convergence of the sequence
$\left\{\mathbf{x}(i)\right\}_{i=0}^{\infty}$ is
$\lambda_{2}\left(\overline{L}\right)>0$. In other words, if
$\lambda_{2}\left(\overline{L}\right)>0$, then there exists an
$\alpha$ such that
$\mathbf{x}(i)\rightarrow\mathbf{x}_{\mbox{\scriptsize{avg}}}
\mbox{a.s.}$ On the contrary, if
$\lambda_{2}\left(\overline{L}\right)=0$ then no choice of
$\alpha$ leads to a.s.~convergence.
\end{theorem}
\begin{proof}
We prove the sufficiency part first. Like Theorem~\ref{theorem-3}
we give a constructive proof. We claim that the choice of
$\alpha=\alpha_{\mbox{\scriptsize{mss}}}=1/2d_{\mbox{\scriptsize{max}}}$
(see eqn.(\ref{th3.6})) leads to a.s.~convergence. To this end,
define the sequence of random variables,
\begin{equation}
\label{th:as1} \mathbf{\xi}_{i} =
\left\|\mathbf{x}(i)-\mathbf{x}_{\mbox{\scriptsize{avg}}}\right\|^{1/2}
\end{equation}
It follows from the properties of finite dimensional real number
sequences (see~\cite{KolFom}) that
\begin{equation}
\label{th:as2}
\mathbf{x}(i)\rightarrow\mathbf{x}_{\mbox{\scriptsize{avg}}}~
\mbox{a.s.}\Leftrightarrow\mathbf{\xi}_{i}\rightarrow 0
~\mbox{a.s.}
\end{equation}
From Theorem~\ref{theorem-3} we note that
\begin{equation}
\label{th:as3} \mathbf{\xi}_{i}\xrightarrow{mss}  0
\end{equation}
Thus $\mathbf{\xi}_{i}\rightarrow 0$ in probability and there exists
a subsequence $\left\{\mathbf{\xi}_{i_{k}}\right\}_{k=0}^{\infty}$
which converges to 0 a.s. Also we note from eqn.(\ref{th3.4}) that
$0\leq\alpha_{\mbox{\scriptsize{mss}}}\leq 1$. Then, from
eqn.(\ref{th3.7}), it follows that
\begin{equation}
\label{th:as4} \rho\left(W-\frac{1}{N}J\right)\leq 1
\end{equation}
Hence from Lemma~\ref{lemma-.4} we have
\begin{eqnarray}
\label{th:as5} \mathbf{\xi}_{i}^{2} & \leq &
\rho\left(W(i-1)-\frac{1}{N}J\right)\mathbf{\xi}_{i-1}^{2}\\
\nonumber & \leq & \mathbf{\xi}_{i-1}^{2}
\end{eqnarray}
Thus $\left\{\mathbf{\xi}_{i}\right\}_{i=0}^{\infty}$ is a
non-increasing sequence of random variables, a subsequence of
which converges a.s.~to 0. By the properties of real valued
sequences $\mathbf{\xi}_{i}\rightarrow 0 ~\mbox{a.s.}$ The
sufficiency part then follows from~(\ref{th:as1}).

The necessary part is trivial, because
$\lambda_{2}(\overline{L})=0$ implies that the network always
separates into at least two components with zero probability of
communication between them. Hence no weight assignment scheme can
lead to a.s.~convergence.
\end{proof}

\textbf{A note on Theorems~\ref{theorem-3} and~\ref{th:as}: } We consider only equal weights,
i.e., all the link weights are assigned the same weight $\alpha$.
However, it is  interesting that, whatever the weights in particular, different weights for different
links, a necessary condition for mss convergence (and a.s.~convergence)
is $\lambda_{2}\left(\overline{L}\right)>0$. This is
because (as argued in Theorem~\ref{th:as}) if
$\lambda_{2}\left(\overline{L}\right)=0$, the network
 separates into two components with zero probability of
communication between each other. Hence, no weight assignment can
lead to mss~convergence. Thus, the necessary condition established
in Theorems~\ref{theorem-3} and~\ref{th:as} for mss~convergence
and a.s.~convergence respectively in the constant link weight case
holds for the more general weight assignments also. In other
words, if we have a weight assignment (with possibly different
weights for different links) for which the consensus algorithm
converges in mss (and a.s.), then we can always find a constant
weight $\alpha$ for which the consensus algorithm converges in mss
(and a.s.)

\section{MSS Convergence Rate} \label{RateofConvergence} We study now the
mss convergence of the algorithm through the
convergence metrics given in
Definitions~\ref{def:mssconvergencefactor}. In the sequel, whenever
we refer to convergence rate of the algorithm, we mean the mss
convergence gain per iteration,
$S_{\mbox{\scriptsize{g}}}(\alpha,\overline{L})$, unless otherwise
stated. We derive bounds on the mss convergence rate of the
algorithm. We assume that
$\lambda_{2}\left(\overline{L}\right)>0$. Hence, by
Theorem~\ref{theorem-3}, there exists $\alpha$, in particular,
$\alpha_{\mbox{\scriptsize mss}}$, leading to mss convergence.
However, given a particular distribution of the Laplacian $L$, the
actual choice of $\alpha$ plays a significant role in determining
the convergence rate. Thus, given a particular
distribution of $L$, we must choose that value of $\alpha$ that
maximizes the convergence speed. From Theorem~\ref{theorem-2}, we
note that, the smaller the
mss-convergence factor $C\left(\alpha,\overline{L}\right)$  given by~(\ref{eq:mssconvergencefactor}) is,  
the faster the convergence is. For a given edge formation
probability distribution~$P$ (and hence $\overline{L}$), the value
of $C\left(\alpha,\overline{L}\right)$ depends on $\alpha$. Thus,
to maximize convergence speed for a given $P$, we perform the
minimization
\begin{eqnarray}
\label{min_conv_func} C^{\ast}\left(\overline{L}\right)& = &
\min_{\alpha}~C(\alpha,\overline{L})\\
\nonumber & = &
\min_{\alpha}~\mbox{E}\left[\rho\left(W-\frac{1}{N}J\right)\right]
\end{eqnarray}
 We present the results in terms of the best achievable mss
convergence rate $S_{\mbox{\scriptsize{g}}}^{\ast}(\overline{L})$
\begin{equation}
 \label{def_Sc}
 S_{\mbox{\scriptsize{g}}}^{\ast}(\overline{L})
=-\ln C^{\ast}(\overline{L})
\end{equation}

The minimization in eqn.~(\ref{min_conv_func}) is difficult.
It depends on the probability
distribution of the Laplacian~$L$. But, by Lemma~\ref{lemma-4},
$C\left(\alpha,\overline{L}\right)$ is convex on~$\alpha$ for a given $\overline{L}$; so, its minimum is
attainable using numerical procedures. In performing this
minimization, we do not need to consider the entire real line for
finding the optimal $\alpha$. The following Lemma provides a range
where the optimal $\alpha$ lies.
\begin{lemma}
\label{lemma-5} Let $\lambda_{2}\left(\overline{L}\right)> 0$.
Then
\begin{equation}
\label{l5.1}
0<\alpha^{\ast}<\frac{2}{\lambda_{N}\left(\overline{L}\right)}
\end{equation}
\end{lemma}
\begin{proof}
Since $\lambda_{2}\left(\overline{L}\right)>0$, by
Theorem~\ref{theorem-3}, we can find $\alpha$ that leads to mss
convergence. But, a necessary condition for mss convergence is
convergence of the mean vector. From section~\ref{Conv_mean}, the
mean converges only if
\begin{equation}
\label{l5.2} 0<\alpha <
\frac{2}{\lambda_{N}\left(\overline{L}\right)}
\end{equation}
Hence, the optimal $\alpha^{\ast}$ leading to fastest mss
convergence must also belong to this range.
\end{proof}
%
We can bound the optimal mss convergence rate
$S_{\mbox{\scriptsize{g}}}^{\ast}(\overline{L})$.
\begin{lemma}
\label{lemma:scstarbound}
If $\lambda_{2}\left(\overline{L}\right)>0$, then
\begin{equation}
\label{Sc_lb}
S_{\mbox{\scriptsize{g}}}^{\ast}(\overline{L})\geq\ln\left(\frac{1}{1-\alpha_{\mbox{\scriptsize
mss}} \mbox{E}\left[\lambda_{2}(L)\right]}\right)
\end{equation}
\end{lemma}
\begin{proof}
By
Theorem~\ref{theorem-3}, if
$\lambda_{2}\left(\overline{L}\right)>0$, then
$\alpha=\alpha_{\mbox{\scriptsize{mss}}}$ leads to mss
convergence and
\begin{eqnarray}
\label{alpha_ms_conv_rate} C\left(\alpha_{\mbox{\scriptsize
mss}},\overline{L}\right)&=&\mbox{E}\left[\rho\left(W-\frac{1}{N}J\right)\right]\\
\nonumber &=& 1-\alpha_{\mbox{\scriptsize
mss}}\mbox{E}\left[\lambda_{2}(L)\right]\\
\label{calfacalfastar} &\geq& C^{\ast}\left(\overline{L}\right)
\end{eqnarray}
The Lemma then follows because
\begin{eqnarray}
\label{Sc_lbbb}
S_{\mbox{\scriptsize{g}}}^{\ast}(\overline{L})&=&\ln\left(\frac{1}{C^{\ast}\left(\overline{L}\right)}\right)\\
\label{Sc_lbb}&\geq&\ln\left(\frac{1}{C\left(\alpha_{\mbox{\scriptsize{mss}}},\overline{L}\right)}\right)\\
\label{Sc_lb_rev}&=&\ln\left(\frac{1}{1-\alpha_{\mbox{\scriptsize
mss}} \mbox{E}\left[\lambda_{2}(L)\right]}\right)
\end{eqnarray}
\end{proof}

\section{Consensus With Communication Constraints: Topology Optimization}
\label{ProbFormCommCost} In the previous sections, we analyzed the
impact of the probability distribution $D$ of the network topology
on the mss convergence rate of the distributed average consensus
algorithm. This section studies the problem of sensor network
topology optimization for fast consensus in the presence of
inter-sensor communication (or infrastructure) cost
constraints. We assume equal link weights  throughout.

We consider $N$ sensors and a symmetric cost matrix $C$, where the
entry $C_{nl}$ is the cost (communication or infrastructure)
incurred per iteration when sensors $n$ and $l$ communicate. The
goal is to design the connectivity graph that leads to the fastest
convergence rate under a constraint on the total communication
cost per iteration. Depending on the structure of the cost matrix
$C$ and the network topology (deterministic or randomized), this
optimization problem may have the following variants:
\begin{enumerate}
\item Fixed topology with equal costs: Here the entries $C_{nl}$
of the cost matrix $C$ are all equal and we look for the optimal
fixed or deterministic topology leading to fastest convergence of
the consensus algorithm. It is easy to see that the equal cost
assumption translates into a constraint on the number of network
links and the optimal solution is essentially the class of
non-bipartite Ramanujan graphs
(see~\cite{Allerton06-K-M,Asilomar06-K-M,tsp06-K-A-M}.)
\item
Fixed topology with different costs (FCCC): In this case the
inter-sensor costs $C_{nl}$ may be different, and we seek the
optimal fixed or deterministic topology leading to fastest
convergence. This is a difficult combinatorial optimization
problem and there is no closed form solution in general.
\item
Random topology with different costs (RCCC): This is the most
general problem, where the costs $C_{nl}$ may be different and we
look for the optimal (random or deterministic) topology leading to
the fastest convergence rate under a communication cost constraint. Because the network is random, it
makes sense to constrain the (network) average (expected)
communication cost per iteration. Likewise, convergence should
also be interpreted in a probabilistic sense, for example, the
mean square convergence. To summarize, in the RCCC problem, we are
concerned with:
\begin{inparaenum}[(i)] \item
 designing the optimal probability of edge formation
matrix $P$, \item under an average communication cost constraint, \item leading
to the fastest mss convergence rate.
\end{inparaenum}
RCCC reduces to  FCCC, if the entries of the optimal $P$ are 0 or
1. In this sense, the RCCC problem relaxes the difficult
combinatorial FCCC problem and, as we will see later, will usually
lead to better overall solutions, especially under medium to low
communication cost constraints. This is because with a fixed
topology, we are forced to use the same network always, while in
the random topology case we can occasionally make use of very good
networks, still satisfying the cost constraint. We can draw an analogy between RCCC  and gossip algorithms
(see~\cite{Boyd-GossipInfTheory}.) However the context and
assumptions of the two problems are different.
Reference~\cite{Boyd-GossipInfTheory} optimizes the gossip probabilities
for a given network topology under the gossip protocol---only two nodes, randomly selected with gossip probability, can communicate at each iteration--- and~\cite{Boyd-GossipInfTheory}  does not impose a communication cost constraint. In contrast, we design the optimal (equal) weight~$\alpha$ and the optimal~$P$ matrix leading to the
fastest mss convergence rate, under an average cost constraint. The topology solution that we determine gives the percentage of time a link is to be used, or, as another interpretation, the probability of error asssociated with reliable communication in a given link. Because signal-to-noise ratio~(SNR) determines often the probability of error, enforcing the topology, i.e., $P$, is like selecting the SNR for each link.
\end{enumerate}

\subsection{Random Topology with Communication Cost Constraints (RCCC)}
\label{ProbFormulation}
We are given $N$
sensors. We model the cost of communication by an
$N\times N$ matrix~$C=C^T$. The entry $C_{nl}\geq 0$, $n\neq l$, is the
cost incurred by a single communication between nodes $n$ and $l$.
Entry $C_{nl}=+\infty$ precludes sensors~$n$ and~$l$ from
communicating. Let $P$ be the probability of edge formation
matrix. The diagonal entries of $P$ are  zero, although each node can access its data with zero
cost. The $P$ matrix induces a probability
distribution on the Laplacian $L(i)$, which at
time~$i$ is a random instantiation based on the~$P$ matrix. The
total cost incurred at stage $i$ is
\begin{eqnarray}
\label{cost_i}  u(i)& =& -\frac{1}{2}\sum_{n\neq
l}L_{nl}(i)C_{nl}\\ \nonumber & = &-\frac{1}{2}\mbox{Tr}(CL(i))
\end{eqnarray}
This follows from  $C$ being symmetric with zero diagonal entries.
Since $L(i)$ is a random matrix, the cost $u_{i}$ incurred at step
$i$ is random. From~(\ref{cost_i}), the expected cost incurred at step~$i$ is
\begin{equation}
\label{exp_cost} \forall
i:\:\:\mbox{E}\left[u_{i}\right]   
 -\frac{1}{2}\mbox{Tr}\left(C\overline{L}\right)
\end{equation}
We consider the distributed averaging consensus model with equal
link weights given in eqns.~(\ref{lin_up_mat}) and~(\ref{cons_W}).
From Section~\ref{MeanSqConv}, mss convergence is determined by
the convergence factor
$C\left(\alpha,\overline{L}\right)=\mbox{E}\left[\rho
\left(W-\frac{1}{N}J\right)\right]$ or the convergence rate
$S_{\mbox{\scriptsize{g}}}(\alpha,\overline{L})$ defined
in~(\ref{def_Sg}). In particular, the smaller
$C\left(\alpha,\overline{L}\right)$ (or larger
$S_{\mbox{\scriptsize{g}}}(\alpha,\overline{L})$) is, the faster
the convergence rate. The expected cost per iteration step in
eqn.~(\ref{exp_cost}) depends on~$\overline{L}$ and hence~$P$,
which are in $1\leftrightarrow 1$ correspondence.

 Let $\mathcal{D}(U)$
be the set of feasible $\overline{L}$ (and hence~$P$) given a
constraint~$U$ on the expected cost per step
\begin{equation}
\label{def_D}
\mathcal{D}(U)=\left\{\overline{L}:-\frac{1}{2}\mbox{Tr}\left(C\overline{L}\right)\leq
U\right\}
\end{equation}
The RCCC problem can then be stated formally as:

{\bf RCCC: Problem formulation.}
\begin{eqnarray}
\label{eqn:rccc-1}
\max_{\alpha,\overline{L}} ~S_{\mbox{\scriptsize{g}}}\left(\alpha,\overline{L}\right)\:\:\:\:\:\:\:\:\:\:\:\:\:\:\:\:\:\:\:\:\:\:\:\:&&\\
\nonumber
\mbox{subject to}\:\:\:\:\:\:\:\:\:\:\:\:\:\:\:\:\:\:\:\:\:\:\:\:\:\:\:\:\:\:
 \overline{L}&=&\overline{L}^{T} \in\mathbb{R}^{N\times N} \\
 \nonumber
 -1\leq &\overline{L}_{nl}&\leq 0,~n,l\in\{1,..,N\},n\neq l \\
 \nonumber
\overline{L}\mathbf{1} &=& \mathbf{0} \\
\nonumber
 -\frac{1}{2}\mbox{Tr}\left(C\overline{L}\right) &\leq& U
\end{eqnarray}
The second inequality constraint comes from the fact that
$\overline{L}_{nl}=-P_{nl},~n\neq l$. The other inequalities
follow from the properties of the Laplacian and the cost
constraint.

\subsection{Alternate Randomized Consensus under Communication Cost Constraints (ARCCC)}
\label{OptCrit} The RCCC problem  in~(\ref{eqn:rccc-1}) is
very difficult to solve. We formulate an alternate randomized consensus under communication cost constraints~(ARCCC) problem. We show successively: \begin{inparaenum}[(i)] \item ARCCC is convex and can be solved by fast numerical optimization procedures; \item
ARCCC is a good approximation to~(\ref{eqn:rccc-1}); and \item ARCCC leads to topologies with good convergence rates. \end{inparaenum}
Point~(i) is  in this section, while points~(ii) and~(iii) are studied in Section~\ref{NumRandCons} where we analyze the performance of ARCCC.

{\bf ARCCC: Problem Formulation.}
\begin{eqnarray}
\label{lemma-7.1}
\max_{\overline{L}} ~\lambda_{2}\left(\overline{L}\right) \:\:\:\:\:\:\:\:\:\:\:\:\:\:\:\:\:\:\:\:\:\:\:\:\:\:\:\:\:\:&&\\
\nonumber
\mbox{subject to}\:\:\:\:\:\:\:\:\:\:\:\:\:\:\:\:\:\:\:\:\:\:\:\:\:\:\:\:\:\:
 \overline{L}&=&\overline{L}^{T} \in\mathbb{R}^{N\times N} \\
 \nonumber
 -1\leq &\overline{L}_{nl}&\leq 0,~n,l\in\{1,..,N\},n\neq l \\
 \nonumber
\overline{L}\mathbf{1} &=& \mathbf{0} \\
\nonumber
 -\frac{1}{2}\mbox{Tr}\left(C\overline{L}\right) &\leq& U
%
%
%
%
\end{eqnarray}

\begin{lemma}
\label{lemma-7} The optimization problem~ARCCC in~(\ref{lemma-7.1})
 is convex.
\end{lemma}
\begin{proof}
From Lemma~\ref{lemma-6}, it follows that the objective
$\lambda_{2}\left(\overline{L}\right)$ is a concave function of
$\overline{L}$. Also, the set of $\overline{L}$ satisfying the
constraints forms a convex set. Hence, ARCCC maximizes a
concave function over a convex set; so, it is convex.
\end{proof}

The optimization problem in Lemma~\ref{lemma-7} is a semidefinite
programming (SDP) problem that can be solved numerically in
efficient ways, see references~\cite{Boyd-CVXBook, SDP-Handbook}
for SDP solving methods (see also~\cite{Boyd-eigopt,
Mesbahi-StateLaplacian} for constrained optimization of graph
Laplacian eigenvalues.)

\section{Topology Optimization: Performance Results}
\label{ARCCC_Perf_Num} In this section,
Subsection~\ref{arcccgoodapproximationtorccc} discusses in what
sense the ARCCC topology optimization problem introduced in
Section~\ref{OptCrit} and eqn.~(\ref{lemma-7.1}) is a good
approximation to the original RCCC topology optimization
formulation of Section~\ref{ProbFormulation} and
eqn.~(\ref{eqn:rccc-1}). Subsection~\ref{AnBounds} establishes
bounds on the optimal value as a function of the communication
constraint. Finally, Subsection~\ref{NumRandCons} illustrates by a
numerical study that the ARCCC optimization obtains topologies for
which the distributed consensus exhibits fast convergence.

\subsection{ARCCC as a Good Approximation to RCCC}
\label{arcccgoodapproximationtorccc}
The difficulty with RCCC stems from the fact that it involves joint optimization over both~$\alpha$ and~$\overline{L}$. For a given $\overline{L}$, there is, in general, no closed
form solution of
\begin{equation}
\label{opt_conv_lambda}
S_{\mbox{\scriptsize{g}}}^{\ast}\left(\overline{L}\right)=\max_{\alpha\in\mathbb{R}}S_{\mbox{\scriptsize{g}}}\left(\alpha,\overline{L}\right)
\end{equation}
We first present a plausible argument of why ARCCC is a good
surrogate for RCCC, and then present numerical results that
justify this argument.

   We present a plausible argument in two steps. First, we replace in RCCC the maximization of $S_{\mbox{\scriptsize{g}}}^{\ast}\left(\overline{L}\right)$ by the maximization of $\mbox{E}\left[\lambda_{2}(L)\right]$. We justify this step by noting that eqn.~(\ref{Sc_lb_rev}) bounds $S_{\mbox{\scriptsize{g}}}^{\ast}(\overline{L})$ from below and this lower bound
shows that larger values of
$\mbox{E}\left[\lambda_{2}(L)\right]$ lead to higher
$S_{\mbox{\scriptsize{g}}}^{\ast}\left(\overline{L}\right)$. This
suggests that, for a given set of distributions
$\overline{L}\in\mathcal{D}(U)$, the quantity
$\mbox{E}\left[\lambda_{2}(L)\right]$ may provide an ordering on
the elements of $\mathcal{D}(U)$ with respect to the mss
convergence rate
$S_{\mbox{\scriptsize{g}}}^{\ast}\left(\overline{L}\right)$.
Hence, a topology with fast convergence rate satisfying the communication constraint~$U$ is provided by the distribution $\overline{L}^{\ast}\in\mathcal{D}(U)$ that
maximizes the quantity $\mbox{E}\left[\lambda_{2}(L)\right]$ over
the set $\mathcal{D}(U)$.

This is not enough to get a reasonable topology optimization problem, since computing $\mbox{E}\left[\lambda_{2}(L)\right]$ is costly, because its evaluation requires costly Monte-Carlo
simulations (see~\cite{ICASSP07-K-M}.) The second step replaces the optimization of $\mbox{E}\left[\lambda_{2}(L)\right]$ by the maximization of $\lambda_{2}(\overline{L})$, which simply involves computing the second eigenvalue of $P=\overline{L}$, no Monte Carlo simulations being involved. This step is justified on the basis of
Lemma~\ref{lemma:lambda2concave}, which upper-bounds
$\mbox{E}\left[\lambda_{2}(L)\right]$ by
$\lambda_{2}(\overline{L})$. This suggests that for
$\mbox{E}\left[\lambda_{2}(L)\right]$ to be large,
$\lambda_{2}(\overline{L})$ should be large.

Putting together the two steps, the RCCC problem in eqn.~(\ref{eqn:rccc-1}) is successively approximated by
\begin{eqnarray}
\label{approx_step} S_{\mbox{\scriptsize{g}}}^{\ast} & = &
\max_{\alpha,\overline{L}\in\mathcal{D}(U)}S_{\mbox{\scriptsize{g}}}(\alpha,\overline{L})\\
\nonumber & \approx &
\max_{\alpha}S_{\mbox{\scriptsize{g}}}(\alpha,\overline{L}^{\ast})\\
\nonumber & = & \widehat{S_{\mbox{\scriptsize{g}}}^{\ast}}
\end{eqnarray}
where $\overline{L}^{\ast}$ is given by
\begin{equation}
\label{max_lambda2L}
\overline{L}^{\ast}=\mbox{arg}~\max_{\overline{L}\in\mathcal{D}(U)}\lambda_{2}\left(\overline{L}\right)
\end{equation}
In general, $\widehat{S_{\mbox{\scriptsize{g}}}^{\ast}}\leq
S_{\mbox{\scriptsize{g}}}^{\ast}$. If
 $S_{\mbox{\scriptsize{g}}}(\alpha,\overline{L})$ was a
non-decreasing function of $\lambda_{2}(\overline{L})$, we would
have $\widehat{S_{\mbox{\scriptsize{g}}}^{\ast}} =
S_{\mbox{\scriptsize{g}}}^{\ast}$.

\begin{figure}[htb]
\begin{center}
\includegraphics[height=2.1in, width=2.5in ]{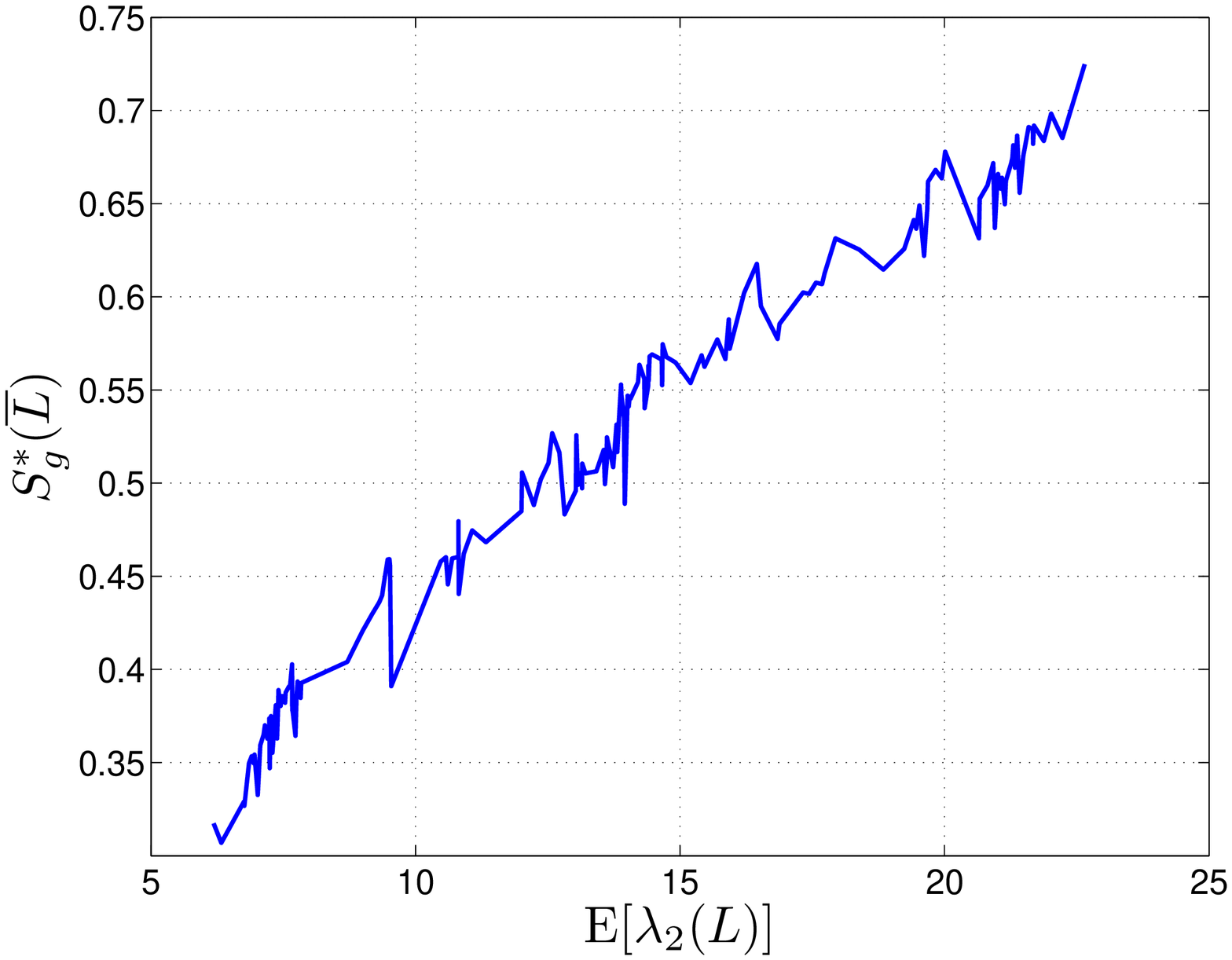}
\includegraphics[height=2.1in, width=2.5in ]{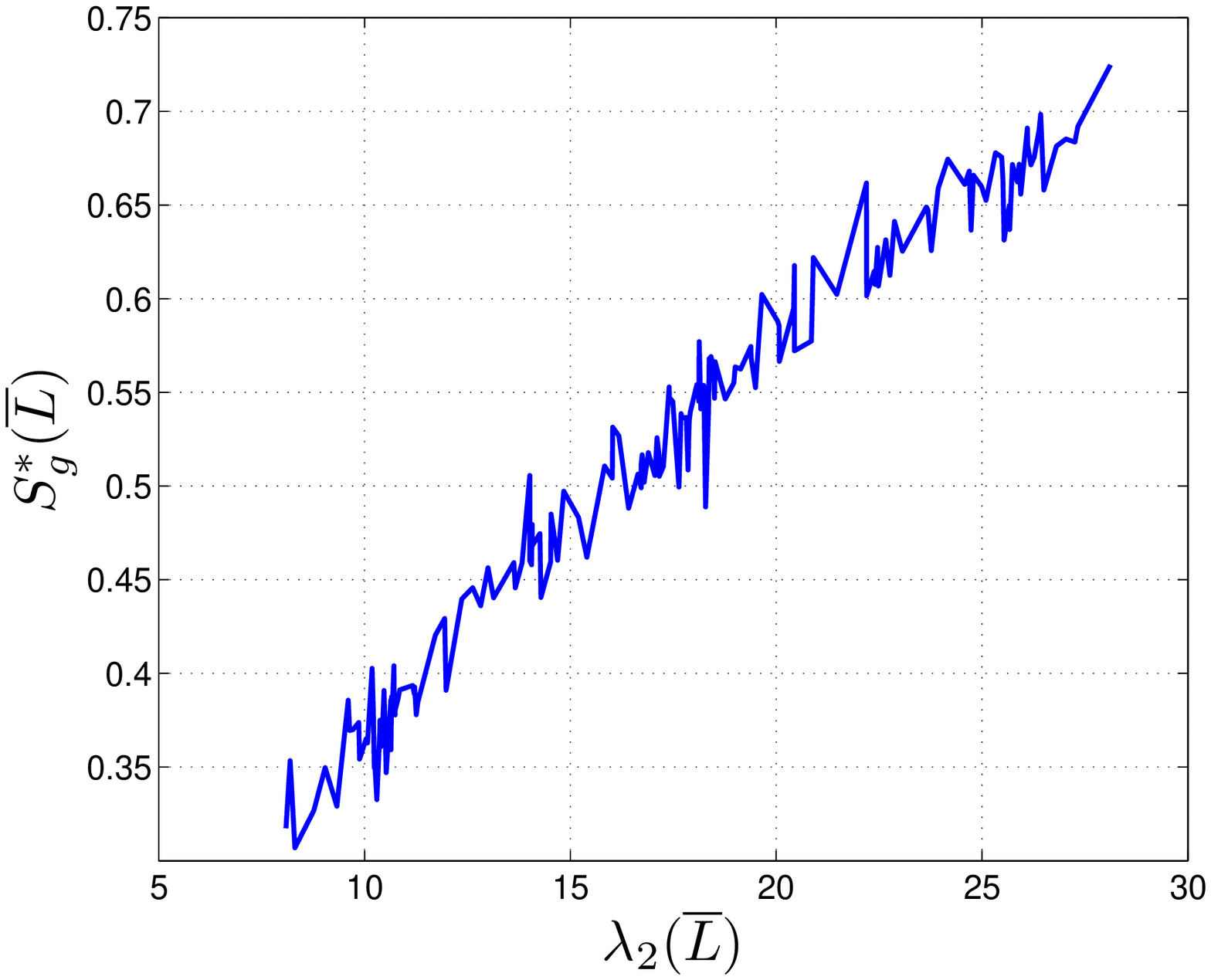}
\caption{Convergence rate
$S_{\mbox{\scriptsize{g}}}^{\ast}\left(\overline{L}\right)$. Left:
with varying $\mbox{E}\left[\lambda_{2}(L)\right]$. Right: with
varying $\lambda_{2}\left(\overline{L}\right)$. The number of
vertices is $N=500$.}
\label{Sc_lambda2_exp}\label{Sc_exp_lambda2}
\end{center}
\end{figure}

We verify by a numerical study how and in what sense
$S_{\mbox{\scriptsize{g}}}^{\ast}\left(\overline{L}\right)$ in~(\ref{opt_conv_lambda}) increases with
$\mbox{E}\left[\lambda_{2}(L)\right]$ and
$\lambda_2\left(\overline{L}\right)$. In our simulation, we choose
a network with $N=500$ sensors and let the average
degree~$d_{\mbox{\scriptsize{avg}}}$ of the network vary in steps
of~$5$ from $10$ to $40$. For each of these~$7$ values
of~$d_{\mbox{\scriptsize{avg}}}$, we construct 200 Erd\"os-R\'enyi
random graphs by choosing at random
$M=d_{\mbox{\scriptsize{avg}}}N/2$ edges of the $N(N-1)/2$
possible pairings of vertices in the network. For each of these
200 random graphs, we generate randomly a probability of formation
matrix~$P$ (hence a probability distribution of $L$) by choosing
for each edge  a weight between~$0$ and~$1$ from a uniform random
distribution. For each such~$P$ matrix, we collect statistics on
the convergence rate
$S_{\mbox{\scriptsize{g}}}^{\ast}\left(\overline{L}\right)$ and
$\mbox{E}\left[\lambda_{2}(L)\right]$ by generating $400$
possible~$L(i)$. For each $P$, we also obtain the corresponding
$\lambda_{2}(\overline{L})$ by eqn.~(\ref{L_bar}). This is an
extensive and computationally expensive simulation.
Fig.~\ref{Sc_lambda2_exp} displays the results by plotting the
convergence rate
$S_{\mbox{\scriptsize{g}}}^{\ast}\left(\overline{L}\right)$ with
respect to $\mbox{E}\left[\lambda_{2}(L)\right]$, left plot, and
with respect to $\lambda_{2}\left(\overline{L}\right)$, right
plot.
 These two plots are remarkably similar and both show that, except for local
oscillations,
  the trend of the convergence
rate $S_{\mbox{\scriptsize{g}}}^{\ast}\left(\overline{L}\right)$
is to increase  with increasing
$\mbox{E}\left[\lambda_{2}(L)\right]$ and
$\lambda_{2}\left(\overline{L}\right)$.  Of course,
 $\lambda_{2}\left(\overline{L}\right)$ is much easier to evaluate
than $\mbox{E}\left[\lambda_{2}(L)\right]$. The plots in
Fig.~\ref{Sc_lambda2_exp} confirm that, given a class
$\mathcal{D}(U)$ of probability distributions of $L$, we can set
an ordering in $\mathcal{D}(U)$ by evaluating the corresponding
$\lambda_{2}\left(\overline{L}\right)$'s, in the sense that a
larger value of $\lambda_{2}\left(\overline{L}\right)$ leads to a
better convergence rate in general (see also~\cite{ICASSP07-K-M},
where part of these results were presented.) This study shows that optimal topologies with respect to ARCCC should be good topologies with respect to RCCC.

%
%
%
%
%
%
%
%
%
%
%
%
%
%
\subsection{ARCCC: Performance Analysis} \label{AnBounds}
 To gain insight into ARCCC, we study the dependence   of the maximum value of its functional
\begin{equation}
\label{def_phi} \phi (U) =
\max_{\overline{L}\in\mathcal{D}(U)}\lambda_{2}\left(\overline{L}\right)
\end{equation}
on the value of the communication cost constraint~$U$. We first establish the concavity of $\phi(U)$.

\begin{lemma}
\label{lemma-8} Given a cost matrix $C$, $\phi(U)$ is a concave
function of $U$.
\end{lemma}
\begin{proof}
Let $0\leq U_{1}\leq U_{2}$ and $0\leq t\leq 1$. Consider the
matrices $\overline{L}^{\ast}_{1}$ and $\overline{L}^{\ast}_{2}$,
such that
\begin{equation}
\nonumber
\lambda_{2}\left(\overline{L}^{\ast}_{1}\right)=\phi\left(U_{1}\right)\:\:
\mbox{and}\:\:
\lambda_{2}\left(\overline{L}^{\ast}_{2}\right)=\phi\left(U_{2}\right)
\end{equation}
It follows that
\[
\overline{L}^{\ast}_{1}\in\mathcal{D}(U_{1})\:\: \mbox{and}\:\:
\overline{L}^{\ast}_{2}\in\mathcal{D}(U_{2})
\]
 Let $\overline{L} =
t\overline{L}^{\ast}_{1}+(1-t)\overline{L}^{\ast}_{2}$.
Then,
\begin{eqnarray}
\label{lemma-8.2}
-\frac{1}{2}\mbox{Tr}\left\{C\overline{L}\right\} & = &
t\left(-\frac{1}{2}\mbox{Tr}\left\{C\overline{L}^{\ast}_{1}\right\}\right)
+
(1-t)\left(-\frac{1}{2}\mbox{Tr}\left\{C\overline{L}^{\ast}_{2}\right\}\right) \\
\nonumber & \leq & tU_{1}+(1-t)U_{2}
\end{eqnarray}
Hence $\overline{L}\in\mathcal{D}\left(tU_{1}+(1-t)U_{2}\right)$.
 From
this we conclude that
\begin{equation}
\label{lemma-8.3} \phi\left(tU_{1}+(1-t)U_{2}\right)\geq
\lambda_{2}\left(\overline{L}\right)
\end{equation}
Now, since $\lambda_{2}\left(\overline{L}\right)$ is a concave
function of $\overline{L}$ (see Lemma~\ref{lemma-6}), we get
\begin{eqnarray}
\label{lemma-8.4} \lambda_{2}\left(\overline{L}\right) & =
&\lambda_{2}\left(t\overline{L}^{\ast}_{1}+(1-t)\overline{L}^{\ast}_{2}\right)
\\
\nonumber & \geq
&t\lambda_{2}\left(\overline{L}^{\ast}_{1}\right)+(1-t)\lambda_{2}\left(\overline{L}^{\ast}_{2}\right)
\\
\nonumber & = & t\phi\left(U_{1})+(1-t)\phi (U_{2}\right)
\end{eqnarray}
Finally, using eqns.(\ref{lemma-8.3} and \ref{lemma-8.4}), we get
\begin{equation}
\label{lemma-8.5} \phi\left(tU_{1}+(1-t)U_{2}\right)\geq
t\phi\left(U_{1}\right)+(1-t)\phi\left(U_{2}\right)
\end{equation}
that establishes the concavity of $\phi(U)$.
\end{proof}
We use the concavity of $\phi(U)$ to derive an upperbound on  $\phi(U)$. Recall that $\mathcal{M}$ is the edge set of the complete graph--the set of all
possible $N(N-1)/2$ edges. Define the set of
realizable edges $\mathcal{E}\subseteq\mathcal{M}$ by
\begin{equation}
\label{def_Esub} \mathcal{E} \left\{(n,l)\in\mathcal{M}~:~C_{nl}<\infty\right\}
\end{equation}
and by $L_{\mathcal{E}}$  the associated Laplacian. Also, let the total cost $C_{\mbox{\scriptsize{tot}}}$
\begin{equation}
\label{def_Ctot} C_{\mbox{\scriptsize{tot}}} = \sum_{(n,l)~\in
~\mathcal{E}}C_{nl}
\end{equation}
The quantity $C_{\mbox{\scriptsize{tot}}}$ is the communication cost per
iteration when all the realizable links are used.

\begin{lemma}
\label{lemma-9}
 Let $C$ be a cost matrix and $U\geq
C_{\mbox{\scriptsize{tot}}}$. Then
$\phi(U)=\lambda_{2}\left(L_{\mathcal{E}}\right)$. If $\mathcal{E} = \mathcal{M}$, then $\phi(U)=N$.
\end{lemma}
\begin{proof}
The best possible case is when all the network links
$(n,l)\in\mathcal{E}$ have probability of formation $P_{nl}=1$
(the links in $\mathcal{E}^{C}$ must have zero probability of
formation to satisfy the cost constraint.) In this case,
$\overline{L} = L_{\mathcal{E}}$. Now, if $U\geq
C_{\mbox{\scriptsize{tot}}}$, then
$L_{\mathcal{E}}\in\mathcal{D}(U)$ and hence the proof follows.
 The
case $\mathcal{E}=\mathcal{M}$ follows from the fact that, for a
complete graph,  $\lambda_{2}\left(L_{\mathcal{M}}\right) = N$
(see~\cite{FanChung, Mohar}.)
\end{proof}
Using the concavity of $\phi (U)$ (Lemma~\ref{lemma-8}), we now
derive a performance bound when $U\leq
C_{\mbox{\scriptsize{tot}}}$.

\begin{lemma}
\label{lemma-10} Let $C$ be a cost matrix. Then
\begin{equation}
\label{lemma-10.1}
 \phi(U)\geq
\left(\frac{U}{C_{\mbox{\scriptsize{tot}}}}\right)\lambda_{2}\left(L_{\mathcal{E}}\right),~~0\leq
U\leq
 C_{\mbox{\scriptsize{tot}}}
\end{equation}
If $\mathcal{E} = \mathcal{M}$, then
\begin{equation}
\label{lemma-10.0}
 \phi(U)\geq \left(\frac{U}{C_{\mbox{\scriptsize{tot}}}}\right)N,~~0\leq U\leq
 C_{\mbox{\scriptsize{tot}}}
\end{equation}
\end{lemma}
\begin{proof}
From Lemma~\ref{lemma-9},
$\phi\left(C_{\mbox{\scriptsize{tot}}}\right)\lambda_{2}\left(L_{\mathcal{E}}\right)$. Then, using the
concavity of $\phi (U)$ (see Lemma~\ref{lemma-8}) and the fact
that $\phi(0)=0$, we have, for $0\leq U\leq
 C_{\mbox{\scriptsize{tot}}}$,
\begin{eqnarray}
\label{lemma-10.2} \phi (U) & = & \phi \left(
\left(\frac{U}{C_{\mbox{\scriptsize{tot}}}}\right)C_{\mbox{\scriptsize{tot}}}\right)
\\ \nonumber & \geq &
\left(\frac{U}{C_{\mbox{\scriptsize{tot}}}}\right)\phi(C_{\mbox{\scriptsize{tot}}})
\\ \nonumber & = &
\left(\frac{U}{C_{\mbox{\scriptsize{tot}}}}\right)\lambda_{2}\left(L_{\mathcal{E}}\right)
\end{eqnarray}
This proves the Lemma. The case $\mathcal{E} = \mathcal{M}$
follows easily.
\end{proof}

Lemma~\ref{lemma-9} states what should be expected, namely: to achieve the optimal performance $\lambda_{2}\left(L_{\mathcal{E}}\right)$ one needs no more than $C_{\mbox{\scriptsize{tot}}}$. Lemma~\ref{lemma-10} is interesting since it states that the ARCCC optimal topology may achieve better performance than the fraction of communication cost it uses  would lead us to expect.  The numerical study in the next Section helps to quantify these qualitative assessments.

\subsection{Numerical Studies: ARCCC}
\label{NumRandCons} 
\begin{figure}[htb]
\begin{center}
\includegraphics[height=2.1in, width=2.5in ]{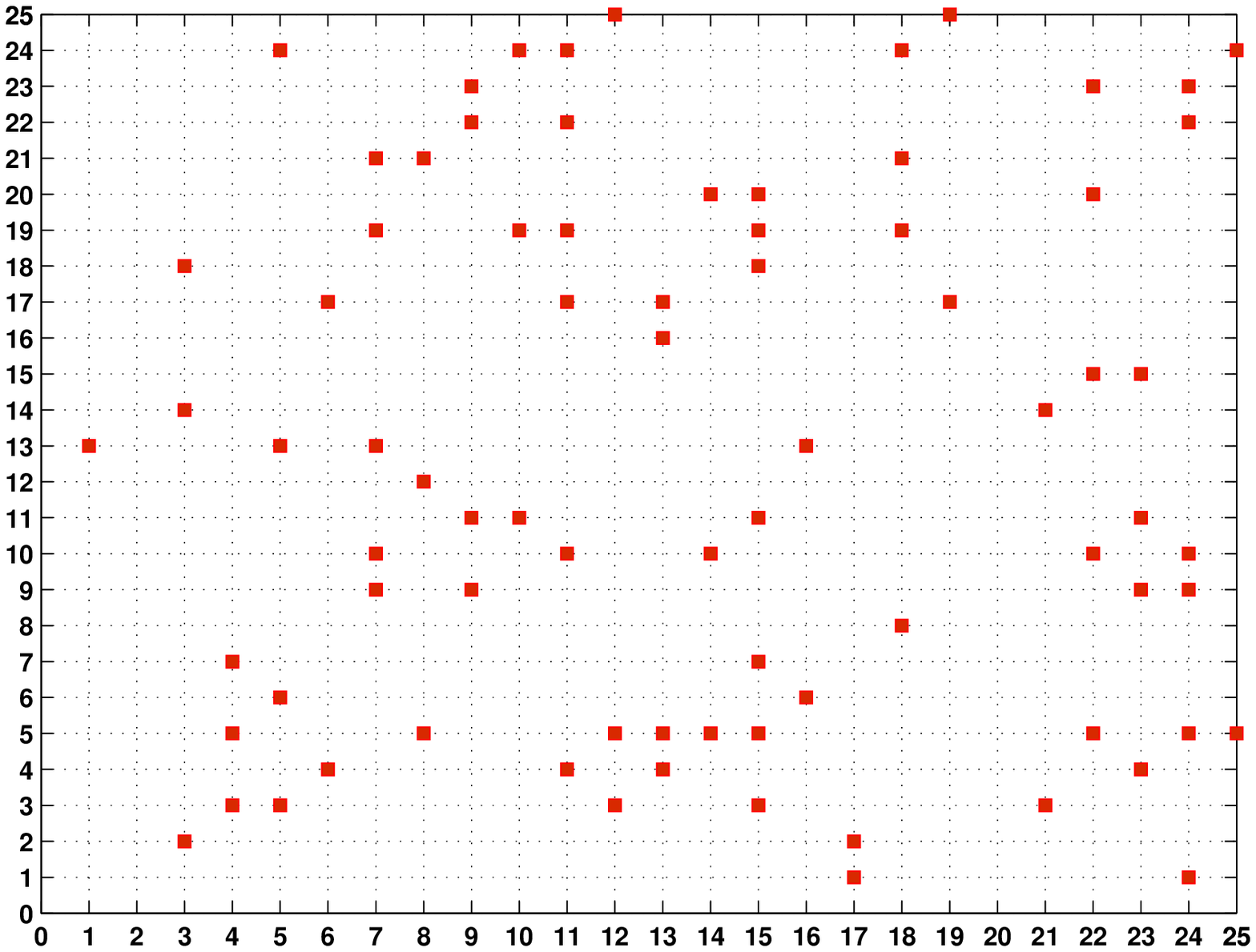}
\includegraphics[height=2.1in, width=2.5in ]{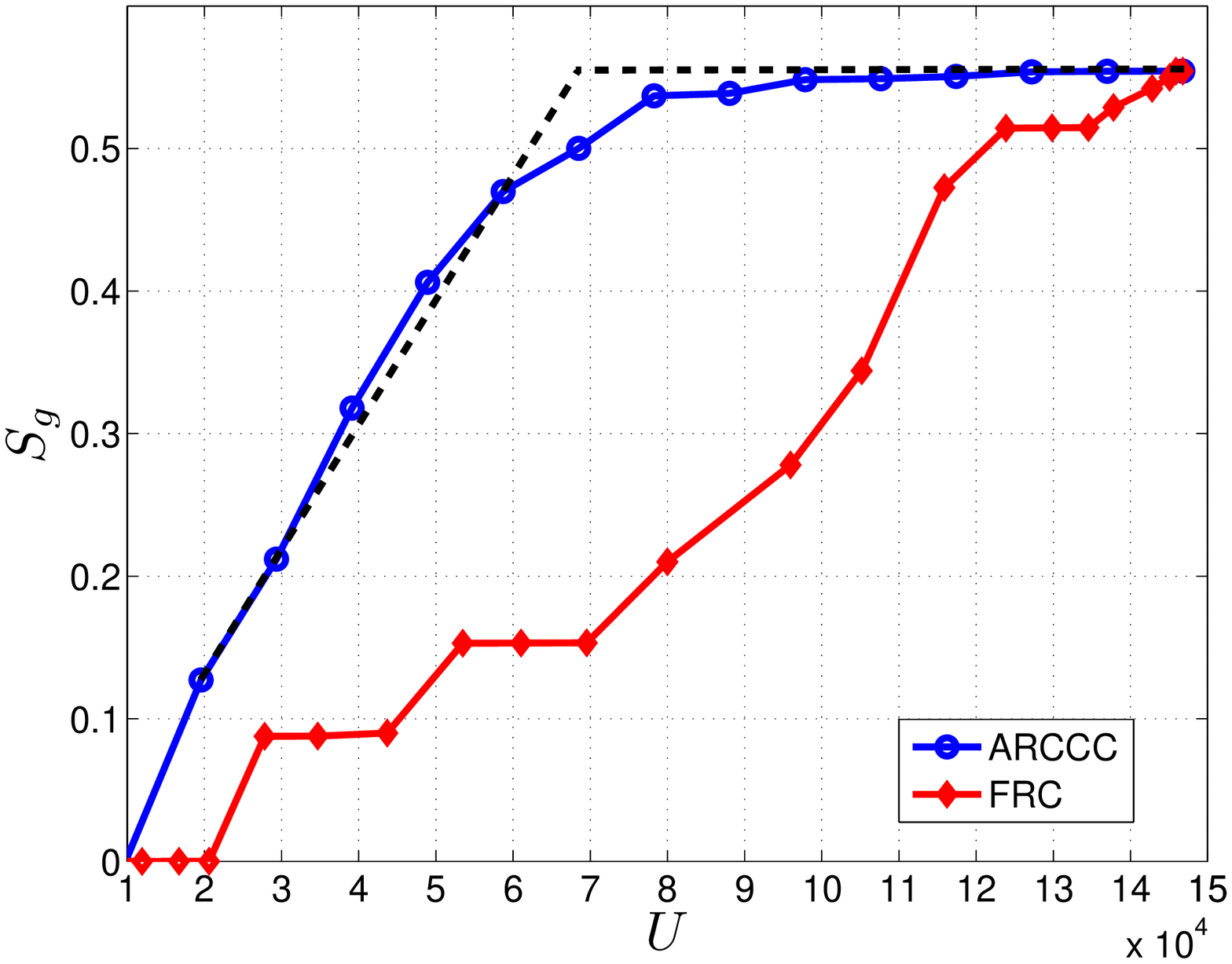}
\caption{Left: Sensor placement of $N=80$ sensors a $25\times 25$
square grid ($\eta = 1$.) Right: Convergence gain
$S_{\mbox{\scriptsize{g}}}$ vs.~communication cost~$U$: ARCCC
optimal topology---top (red) line; FRC topology---bottom (blue)
line.} \label{SensPlacement} \label{ARCCCvsNNC}
\end{center}
\end{figure}

This Section solves the ARCCC semidefinite
programming optimization  given by~(\ref{lemma-7.1}). It solves for~$P$, which assigns to each realizable link its probability of error (aka, SNR), or the fraction of time it is expected to be active. We compare
the ARCCC optimal topology to a fixed radius connectivity~(FRC)
topology detailed below. The sensor network is displayed on the
left of Fig.~\ref{SensPlacement}. We deploy $N=80$ sensors
uniformly on a $25\times 25$ square grid on the plane. The set
$\mathcal{E}$ of realizable links is constructed by choosing
$|\mathcal{E}|=9N$ edges randomly from the set $\mathcal{M}$ of
all possible edges. We assume a geometric propagation model: the
communication cost is proportional to the square of the Euclidean
distance $d_{nl}$ between sensors~$n$ and~$l$ 
\begin{equation}
\label{def_cmat} C_{nl} = \left\{\begin{array}{ll}
                            \eta d_{nl}^{2} & \mbox{if
                            $(n,l)\in\mathcal{E}$}\\
                            \infty & \mbox{otherwise}
                            \end{array}\right.
\end{equation}
where $\eta$ is an appropriately chosen constant.   With the FRC
network, a sensor $n$ communicates with all other sensors $l$
($C_{nl}<\infty$) that lie within a radius~$R$. The FRC topology
is an instantiation of a fixed, i.e., not random, topology with a
fixed cost incurred per iteration.

%

%

Fig.~\ref{ARCCCvsNNC} on the right plots, as a function of the cost constraint~$U$, the per step convergence
gain
$S_{\mbox{\scriptsize{g}}}=\widehat{S_{\mbox{\scriptsize{g}}}^{\ast}}$
for the ARCCC optimal topology (top blue line) and the per step convergence
gain $S_{\mbox{\scriptsize{g}}}$ of the FRC topology (bottom red line).  The ARCCC optimal topology converges much faster than the FRC topology, with the improvement being
more significant at medium to lower values of $U$.

The ARCCC topology has a markedly nonlinear behavior, with two
asymptotes: for small~$U$, the sharp increasing asymptote,  and the  asymptotic horizontal asymptote (when all the
realizable edges in $\mathcal{E}$ are used.) The two meet at the
knee of the curve $\left(U=6.9\times
10^{4},S_{\mbox{\scriptsize{g}}}=.555\right)$. For $U=6.9\times10^4$, the
ARCCC convergence rate is
$\widehat{S_{\mbox{\scriptsize{g}}}}=.505$, while FRC's is
$S_{\mbox{\scriptsize{g}}}=.152$, showing that ARCCC's topology is
$3.3$~times faster than FRC's. For this example, we compute
$C_{\mbox{\scriptsize{tot}}}=14.7\times 10^{4}$, which shows that
ARCCC's optimal topology achieves the asymptotic performance while
using less than 50~\% of the communication cost.


\section{Conclusions}
\label{Conclusions} The paper  presents the design of the topology
of a sensor network to maximize the convergence rate of the
consensus algorithm as a convex optimization problem. We consider
that the communication channels among sensors may fail at random
times, that communication among sensors incurs a cost, and that
there is an overall communication cost constraint in the network.
We first establish necessary and sufficient conditions for mss
convergence and a.s.~convergence in terms of the expected value of
the algebraic connectivity of the random graph defining the
network topology and in terms of the algebraic connectivity of the
average topology.  We apply semidefinite programming to solve
numerically for the optimal topology design of the random network
subject to the communication cost constraint. 
Because the topology is random, the solution to this optimization specifies for each realizable link its probability  of error (aka, SNR), or the fraction of time the link is expected to be active.
We show by a
simulation study that the resulting topology design can improve by
about 300~\% the convergence speed of average consensus over more
common designs, e.g., geometric topologies where sensors
communicate with sensors within a fixed distance. Our study also
shows that the optimal random topology can achieve the convergence
speed of a non-random network at a fraction of the cost.

\bibliographystyle{IEEEtran}
\bibliography{IEEEabrv,BibOptEqWeights}

\end{document}